\documentclass[conference]{IEEEtran}
\usepackage{graphicx}
\usepackage{float}
\usepackage[caption=false]{subfig}
\usepackage[space]{grffile}
\usepackage{latexsym}
\usepackage{enumerate}
\usepackage{textcomp}
\usepackage{amsfonts,amsmath,amssymb}
\usepackage{bm}

\usepackage{url}
\usepackage{algorithm}
\usepackage{algpseudocode}
\usepackage{hyperref}
\usepackage{tabularx}
\usepackage{longtable}
\usepackage{multirow,booktabs,makecell}
\usepackage{xcolor}
\usepackage{diagbox}
\usepackage{siunitx}
\usepackage{cleveref}

\usepackage{caption}
\newcommand{\Jnees}[0]{J_{NEES}}

\def\mr[#1]#2#3{\multirowcell{#2}[#1]{#3}}

\newcommand{\Ignore}[1]{}

\newcommand{\kLst}{k-1}
\newcommand{\kCur}{k}

\newcommand{\x}[1]{\mathbf{x}_{#1}}
\newcommand{\z}[1]{\mathbf{z}_{#1}}

\newcommand{\uVec}[1]{\mathbf{u}_{#1}}
\newcommand{\vVec}[1]{\mathbf{v}_{#1}}
\newcommand{\wVec}[1]{\mathbf{w}_{#1}}
\newcommand{\F}[1]{\mathbf{F}_{#1}}
\newcommand{\Ft}[1]{\mathbf{F}_{#1}^\top}
\newcommand{\HM}[1]{\mathbf{H}_{#1}}
\newcommand{\HMt}[1]{\mathbf{H}_{#1}^\top}
\newcommand{\Q}[1]{\mathbf{Q}_{#1}}

\newcommand{\R}[1]{\mathbf{R}_{#1}}

\newcommand{\ex}[1]{\mathbf{e}_{\mathbf{x},#1}}
\newcommand{\ez}[1]{\mathbf{e}_{\mathbf{z},#1}}
\newcommand{\nees}[1]{\epsilon_{\mathbf{x},#1}}
\newcommand{\nis}[1]{\epsilon_{\mathbf{z},#1}}
\newcommand{\avgnees}[1]{\bar{\epsilon}_{\mathbf{x},#1}}
\newcommand{\avgnis}[1]{\bar{\epsilon}_{\mathbf{z},#1}}
\newcommand{\W}[0]{\mathbf{W}}
\newcommand{\V}[0]{\mathbf{V}}

\newcommand{\E}[1]{\mathrm{E}\left[#1\right]}

\newcommand{\xCond}[2]{\hat{\mathbf{x}}_{#1|#2}}

\newcommand{\covCond}[2]{\mathbf{P}_{#1|#2}}
\newcommand{\covCondi}[2]{\mathbf{P}^{-1}_{#1|#2}}
\newcommand{\covCondTrue}[2]{\mathbf{P}^a_{#1|#2}}

\newcommand{\zCond}[2]{\hat{\mathbf{z}}_{#1|#2}}

\newcommand{\innovCov}[2]{\mathbf{S}_{#1|#2}}
\newcommand{\nx}[0]{n_{\mathbf{x}}}
\newcommand{\nz}[0]{n_{\mathbf{z}}}
\newcommand{\Kw}[1]{\mathbf{K}_{#1}}
\newcommand{\Kwt}[1]{\mathbf{K}_{#1}^\top}
\newcommand{\KK}[1]{\mathbf{K}_{#1}}
\newcommand{\KKt}[1]{\mathbf{K}_{#1}^\top}

\newcommand{\Snu}[1]{\mathbf{S}_{#1}}

\begin{document}

\title{Time Dependence in Kalman Filter Tuning}

\author{
\IEEEauthorblockN{Zhaozhong Chen, Christoffer Heckman}
\IEEEauthorblockA{Department of Computer Science \\ University of Colorado Boulder \\
430 UCB \\ Boulder, CO 80309 \\ email: Christoffer.Heckman@colorado.edu}
\and
\IEEEauthorblockN{Simon Julier}
\IEEEauthorblockA{Department of Computer Science \\ University College London\\
66--72 Gower Street, \\ London WC1E 6BT, UK \\ email: s.julier@ucl.ac.uk}
\and
\IEEEauthorblockN{Nisar Ahmed}
\IEEEauthorblockA{Smead Aerospace Engineering Sciences \\ University of Colorado Boulder \\
429 UCB \\ Boulder, CO 80309 \\ email: Nisar.Ahmed@colorado.edu}
}
\maketitle

% \twocolumn[{%
% \renewcommand\twocolumn[1][]{#1}%
% \maketitle
% \begin{center}
%     \label{fig:first_img}
%     \centering
%     \captionsetup{type=figure}
%     %,height=4.5cm
%     \includegraphics[width=.31\textwidth]{figs/dt0105_tracking1d_page1.eps} \ \ \
%     \includegraphics[width=.31\textwidth]{figs/dt01_tracking1d_page1.eps} \ \ \
%     \includegraphics[width=.31\textwidth]{figs/dt05_tracking1d_page1.eps}
%     \captionof{figure}{We want to tune the noise parameters $v$, $w$ for a stochastic system by searching the minimum cost (Section \ref{sct:new_measure}) on the surface. Our proposed method use multiple sample time $dt = 0.1$ and $dt = 0.5$ and cost surface has a more clear minimum. Thus, we can achieve a more precise estimation than using a single sample time ($dt = 0.1$ or $dt = 0.5$). Red circle: Groundtruth, Black diamond: Possible estimation results from multiple optimizations.}
% \end{center}%
% }]

\maketitle

\begin{abstract}

In this paper, we propose an approach to address the problems with ambiguity in tuning the process and observation noises for a discrete-time linear Kalman filter. Conventional approaches to tuning (e.g. using normalized estimation error squared and covariance minimization) compute empirical measures of filter performance and the parameter are selected manually or selected using some kind of optimization algorithm to maximize these measures of performance. However, there are two challenges with this approach. First, in theory, many of these measures do not guarantee a unique solution due to observability issues. Second, in practice, empirically computed statistical quantities can be very noisy due to a finite number of samples. We propose a method to overcome these limitations. Our method has two main parts to it. The first is to ensure that the tuning problem has a single unique solution. We achieve this by simultaneously tuning the filter over multiple different prediction intervals. Although this yields a unique solution, practical issues (such as sampling noise) mean that it cannot be directly applied. Therefore, we use Bayesian Optimization. This technique handles noisy data and the local minima that it introduces. We demonstrate our results in a reference example and demonstrate that we are able to obtain good results. We share the source code for the benefit of the community\footnote{\url{https://github.com/arpg/kf_bayesopt}}.
\end{abstract}

%\begin{IEEEkeywords}
%Kalman filter, filter tuning, Bayesian optimization, nonparametric regression, machine learning.
%\end{IEEEkeywords}

%\IEEEpeerreviewmaketitle

\section{Introduction}
State estimation through Kalman filters consists of two main steps: \emph{state prediction} followed by a \emph{measurement update}, both predicated on models of the system. The state prediction step uses a process model to predict how the state evolves over time. The measurement update step uses an observation model to relate a measured quantity to the state estimate. Since both the process and observation models are imperfect, errors in these models are treated as random noise terms that are injected into the system. Most designs assume the noise in these systems is white, zero mean and uncorrelated. As a result, filter tuning consists of choosing the values of the process and observation noise covariances, thereby fully defining the noise distribution.

Given the critical role that tuning plays in the performance of these algorithms, multiple techniques for tuning filters have been developed \cite{aakesson2007tool,AKESSON2008769, powell2002automated}. Perhaps the simplest approach is to use a two-stage divide-and-conquer strategy. In the first stage, the observation covariance is estimated by operating the system in lab conditions and monitoring the sensor noise characteristics. In the second stage, the observation covariance is held fixed, and the process noise covariance is determined. Since the process noises contain information about the state disturbances and dynamic model uncertainties, which often cannot be reproduced in lab settings, the covariance is often chosen by collecting data from an operational domain and quantifying the quality of the estimates. Typically a performance cost is assigned, and the process noise covariance is adjusted to minimize the value of that cost.

Other approaches include `black box' auto-tuning methods \cite{chen2018weak, mu2009automatic, scardua2016automatic}, which construct a cost function to be minimized based on properties of the state or statistical principles regarding estimates produced. We demonstrate that, even in simple examples, these methods do not guarantee convergence to a unique optimum, and frequently converge to the incorrect optimum. We also shed light on the relationship between noise parameter identifiability and use of consistency metrics as fitness measures for auto-tuning methods, particularly to understand how mismatches between the filter-assumed and true system noise parameters impacts search algorithm convergence. Novel solutions to these issues are presented via measurement and process noise perturbation strategies, and demonstrated on reference examples via Bayesian optimization.

\section{Prelminaries}
\label{sct:preliminaries}

\subsection{Discrete and Continuous Time Systems}

Our approach depends upon adjusting the prediction interval in the Kalman filter. Therefore, it is important to understand the relationship between the discrete and continuous time systems. The state of the system at time $t$ is $\x{t}$. The system is described by continuous time process model and observation models, %These are
\begin{equation}
\label{eq:con_sys_model}
\begin{aligned}
\dot{\mathbf{x}}_t &= \mathbf{A}_t \x{t} + \mathbf{G}_t \uVec{t} + \boldsymbol{\Gamma}_t \vVec{t}, \\
\z{t} &= \mathbf{H}_t \x{t} + \wVec{t},
\end{aligned}
\end{equation}
where $\uVec{t}$ is the control input, the process noise is the additive white process $\vVec{t}$ with intensity $\V$, and the measurement noise is an additive white noise process $\wVec{t}$ with continuous time intensity $\W$. 
%
%Most systems are implemented in discrete time. Therefore, 
In discrete time, the state at timestep $k$ is $\x{k}$. The system evolution from timestep $\kLst$ to $\kCur$ is
\begin{equation}
\x{\kCur}=\F{\kCur}\x{\kLst}+\mathbf{B}_{\kCur}\uVec{\kCur}+\vVec{\kCur},
\label{eq:dynModel}
\end{equation}
\noindent where $\uVec{\kCur}$ is the control input and $\vVec{\kCur}$ is the process noise, which is assumed to be zero
mean and independent with covariance $\Q{\kCur}$. The observation model is
\begin{equation}
\z{\kCur}=\HM{\kCur}\x{\kCur}+\wVec{\kCur}, \label{eq:measModel}
\end{equation}
\noindent where $\wVec{\kCur}$ is the observation noise.

The discrete-time system is derived from the continuous time system using techniques such as Van Loan's method~\cite{mohindergrewalVanLoanMethod2015},
\begin{equation}
\begin{aligned}
&\F{k} = 
e^{\mathbf{A}_t\Delta t}, \ \
\mathbf{B}_k = 
\int_{0}^{\Delta t} e^{\mathbf{A}_t m} \mathrm{d} m,\\
&\Q{k} =  \int_{0}^{\Delta t} e^{\mathbf{A}_tm} \boldsymbol{\Gamma} \V \boldsymbol{\Gamma}^T e^{\mathbf{A}^Tm} \mathrm{d} m .
\label{eq:VanLoans}
\end{aligned}
\end{equation}
%Other approachs, such as ``sample and hold'' can be used to approximate these quantities. but these lead to difficulties. %\sjj{Not sure what the standard reference is for this --- and whether we actually need it or not.}
If the observation is from an integrating sensor, the discrete time observation vector is $\R{k} =  \W / \Delta t$ \cite{mohindergrewalVanLoanMethod2015}.
%\begin{equation}
%\R{k} =  \frac{\W}{\Delta t}.
%\end{equation}
For a non-integrating sensor $\R{k}=\R{t}$, i.e.\ it is independent of $\Delta{t}$.

\subsection{Kalman Filter}

A Kalman filter can be used to find the optimal state estimate \cite{Kalman-JBE-1961}, via a two stage process of prediction followed by measurement update. The prediction is
\begin{align}
\xCond{\kCur}{\kLst}&=\F{\kCur}\xCond{\kLst}{\kLst} +\mathbf{B}_{\kCur}\uVec{\kCur} \label{eq:kalman1}\\
\covCond{\kCur}{\kLst}&=\F{\kCur}\covCond{\kLst}{\kLst}\Ft{\kCur}+\Q{\kCur}
\end{align}
\noindent while the update is
\begin{align}
\xCond{\kCur}{\kCur}&=\xCond{\kCur}{\kLst}+\Kw{\kCur}\ez{\kCur},\\
\covCond{\kCur}{\kCur}&=\covCond{\kCur}{\kLst}-\Kw{\kCur}\Snu{\kCur|\kLst}\Kwt{\kCur},\\
\Snu{\kCur|\kLst}&=\HM{\kCur}\covCond{\kCur}{\kLst}\HMt{\kCur}+\R{\kCur}\\
\Kw{\kCur}&=\covCond{\kCur}{\kLst}\HMt{\kCur}\Snu{\kCur|\kLst}^{-1} \label{eq:kalman6}
\end{align}
%Many other recursive estimators, including particle filters and grids have a similar two-step structure.

%\sjj{I started reformulating the problem to estimate $V$ and $W$. The logic is that gives us a description that's invariant with respect to the timestep length. I haven't followed this through consistently though.}

One important issue with this method is tuning: given $\F{\kCur}$ and $\HM{\kCur}$, the process and observation noise processes $\V$ and $\W$ must be determined. This is normally achieved by exploring different values of $\V$ and $\W$ and applying a fitness measure.

\subsection{Parameter Fitness and Tuning}

Two widely used measures for fitness are the \emph{normalized estimation error squared
(NEES)} and the \emph{normalized innovation error squared (NIS)}. The NEES and NIS is
computed from

\begin{align}
&\nees{k} = \ex{k}^T \covCond{k}{k}^{-1} \ex{k}, \label{eq:neesDef}\\
&\nis{k} = \ez{k}^T \innovCov{k}{k-1}^{-1} \ez{k}, \label{eq:nisDef}
\end{align}

where $\ex{k} = \x{k} - \xCond{k}{k}, \ez{\kCur}= \z{\kCur} - \zCond{\kCur}{\kLst}$.
%because we need to know $\x{k}$, NEES requires a groundtruth measurement of the system state. The NIS, on the other hand, only depends on the observation sequence and does not require knowledge of groundtruth. It is computed from
%\begin{equation}
%\nis{k} = \ez{k}^T \innovCov{k}{k-1}^{-1} \ez{k}, \label{eq:nisDef}
%\end{equation}
%where $\ez{k}$ is the innovation vector. 
If the filter is statistically consistent, it can be shown that the expected values of the NEES and the NIS are~\cite{Bar-Shalom2001}
\begin{equation}
\begin{aligned}
    \E{\nees{k}}{\approx}\nx, \ \ 
    \E{\nis{k}}{\approx}\nz,
\end{aligned}
\label{eqn:nees_nis}
\end{equation}
Although the $\nis{k}$ and $\nees{k}$ are widely used, they have the property that they are bounded from below (by 0) but not from above. This naturally introduces a bias or asymmetry in the measure. To overcome this, we use a log measure instead:
\begin{equation}
    \begin{aligned}
    &\Jnees =   \left | \log \left(\frac{\sum_{k=1}^{T}{\avgnees{k}}/T}{\nx} \right) \right|, \\
    &\avgnees{k} = \frac{1}{N} \sum_{i=1}^{N}{\nees{k}^i}.
    \end{aligned}
    \label{eqn:jnees}
\end{equation}
\noindent where $N$ is the number of Monte Carlo runs and $T$ is the period of sampling. $\Jnees$ is not bounded. However, when the filter is consistent, $\Jnees=0$.

%relatedwork.tex
\subsection{Related Work}
%%\nra{might actually want to put this section after math preliminaries...}
Though the problem of Kalman filter tuning has been widely studied, it remains a challenging open problem for which no single best technique exists \cite{zhangIdentificationNoiseCovariances2020,dunik2017noise}. 
%Many tuning approaches consider analytically grounded parameter estimation strategies based on maximum likelihood or approximate nested 
These include: maximum likelihood and
Bayesian inference \cite{Bishop2006}, least squares for data processed via Kalman smoothing \cite{barratt2020fitting}, and auto-/cross-correlation analysis \cite{dunik2020covariance}. These methods are theoretically advantageous for well-defined linear systems where noise models have known structure, and are useful in online settings. Yet, they can also suffer from numerical stability and implementation issues, making them harder to use. Moreover, they are difficult to generalize for non-linear filters, e.g. since the optimal set of noise parameters in linearization-based filters can vary significantly with system state and time \cite{ko2009gp}. 
%%have certain advantages, such as...\nra{e.g. ability to exploit problem structure for well defined systems or noise models...}, they are not necessarily well suited to certain kinds of problems \nra{which?...and also have numerical stability/implementation issues...makes them harder to use}

%The family of `black box' optimization approaches considered here offers a computationally attractive and flexible alternative, whereby filter-assumed noise parameters are adjusted via search algorithms to maximize a set of filter output fitness measures, which are assessed on candidate filter runs against truth model simulations and/or recorded measurement logs. In addition to being highly parallelizable in most cases, black box optimization can readily leverage useful but complex stochastic fitness measures that do not yield tractable `well-behaved' expressions for objective functions and gradients with respect to unknown noise parameters. %(e.g. as the underlying cost surface may be non-smooth and discontinuous). 

The family of `black box' optimization approaches considered here are widely used. The defining features of black box methods are the choice of filter output fitness measure and search algorithm. Powell \cite{powell2002automated} proposed using a mean weighted filter state error norm as a fitness measure to be minimized via downhill simplex search. In earlier work, Oshman and Shaviv \cite{oshman2000optimal} presented a fitness measure based on chi-square tests for NEES consistency (evaluated using truth model simulations) to tune process noise covariance parameters via genetic algorithms. More recently, \cite{chen2018weak} developed a technique using Bayesian optimization search and generalized filter output fitness measures based on NIS consistency tests with real/logged data, as well as NEES consistency tests with truth model simulation runs. Other metrics closely related to NIS consistency assessment \cite{saha2013robustness, piche2016online, gibbs2013new} could also be adapted as fitness measures. 

%An important question for black box approaches concerns their convergence behavior for different fitness metrics.
While search methods like genetic algorithms and Bayesian optimization can explore the global parameter space, the observability (i.e. identifiability) of noise parameters relative to estimation error and consistency-based fitness metrics is not well understood. For instance, \cite{oshman2000optimal} noted that their approach generally converged towards an infinite basin of feasible parameters which all satisfy the NEES consistency criterion, without necessarily minimizing the resulting steady state $\mathbf{P}$. As such, \cite{oshman2000optimal} also proposed a fitness measure to minimize filter covariance, while ensuring NEES consistency within some tolerance. However, the general conditions for convergence toward unique or multiple/infinite solutions remain unclear. Ref. \cite{zhangIdentificationNoiseCovariances2020} addresses the observability of $\mathbf{Q}$ and $\mathbf{R}$ in discrete time Gauss-Markov linear systems by deriving a matrix rank test. This is theoretically useful for assessing uniqueness of time invariant $\mathbf{Q}$ and $\mathbf{R}$ parameters, provided the hypothesized matrix structures match the true system behavior. Otherwise, the correctness and sensitivity of the matrix structures and values cannot be readily deduced. 

\section{The Problem of Observability}
\label{sct:observability}

The non-uniqueness (non-observability) of noise parameters via consistency-based fitness metrics is a key problem for black box tuning approaches. 
%~\cite{zhangIdentificationNoiseCovariances2020}.
We illustrate this using the following linear example. We seek to tune the process and observation noise processes for a 1D particle. The particle's state is its position and velocity,
\begin{equation*}
 \x{t}=\begin{bmatrix}
 x_t&\dot{x}_t
 \end{bmatrix}^\top.
\end{equation*}
It moves with a constant velocity with noise injected into the acceleration. The particle's position is periodically observed by a non-integrating sensor. Therefore, the continuous time equations are
\begin{align*}
\mathbf{A} =
\begin{bmatrix}
0 & 1 \\
0 & 0 
\end{bmatrix}, \ \ 
\mathbf{G} =
\begin{bmatrix}
0 \\
1  
\end{bmatrix}, \ 
\HM{} =
\begin{bmatrix}
1 & 0  
\end{bmatrix}, \ 
\boldsymbol{\Gamma} = 
\begin{bmatrix}
0 \\ 
1  
\end{bmatrix}. \ 
\end{align*}
Van Loan's method yields the familiar discrete-time equations
\begin{equation}
\begin{aligned}
&\F{k} = 
\begin{bmatrix}
1 & \Delta t \\
0 & 1
\end{bmatrix}, \ \
\mathbf{B}_k = 
\begin{bmatrix}
\Delta t^2/2 \\
\Delta t
\end{bmatrix}, \ \ 
\HM{k} = 
\begin{bmatrix}
1 & 0
\end{bmatrix},\\
&\Q{k} = \V\begin{bmatrix}
 \Delta t^3/3 & \Delta t^2/2 \\
 \Delta t^2/2 & \Delta t
\end{bmatrix}, \ \ \R{k} = \W.
\end{aligned}
\label{eq:dis_1d_sys_model}
\end{equation}

\begin{figure}
\begin{center}
\includegraphics[width=0.8\linewidth]{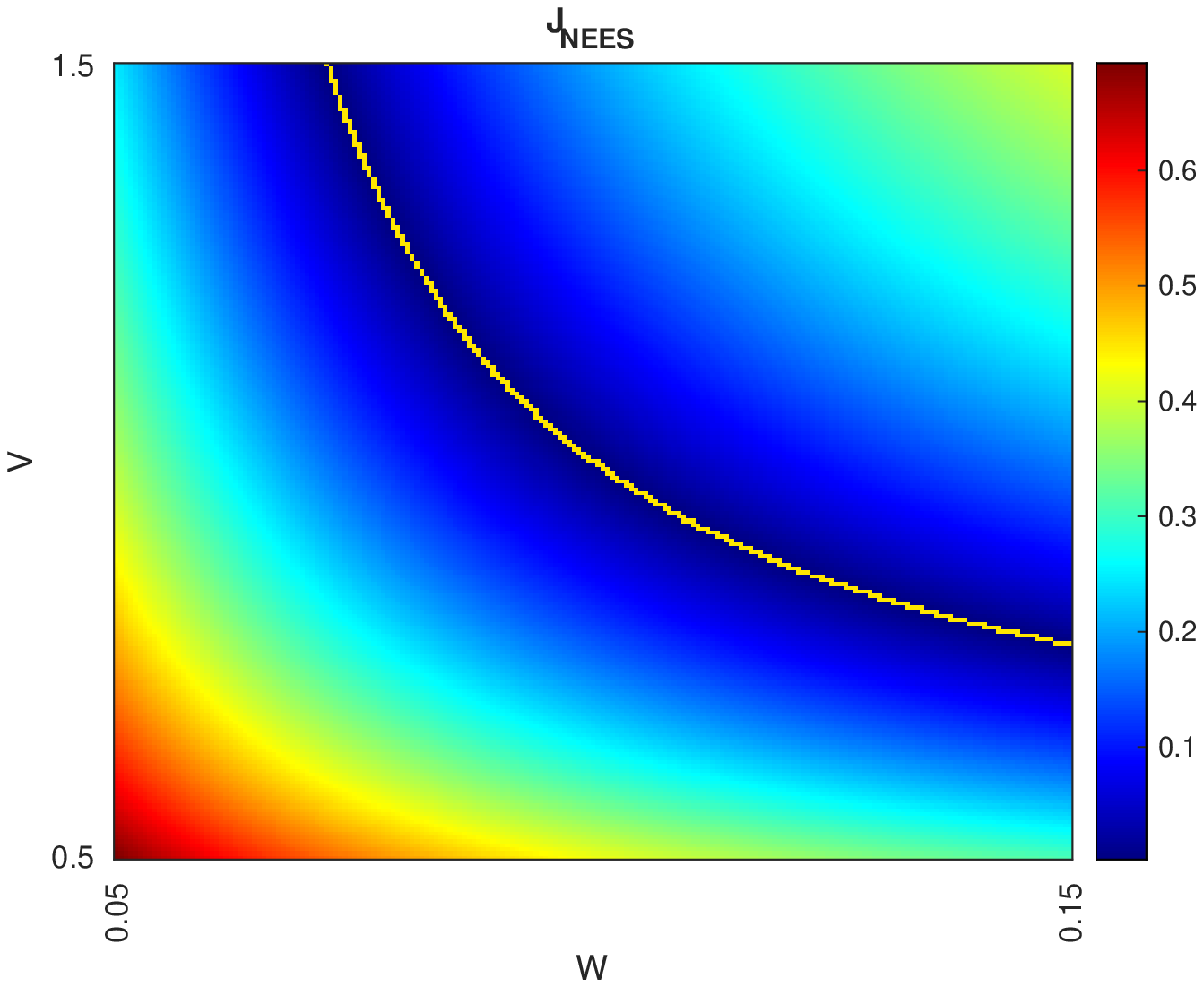}
\end{center}
\caption{Theoretically computed $\Jnees(\V,\W,\V^a,\W^a)$ for different values of $\V$ and $\W$.}
\label{fig:scan_results}
\end{figure}

\begin{figure}
\begin{center}
\includegraphics[width=0.8\linewidth]{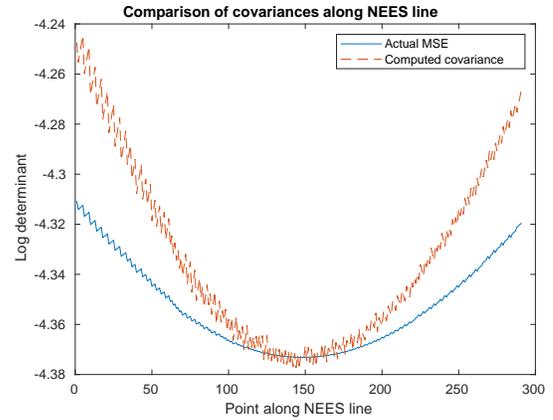}
\end{center}
\caption{Log determinant of the actual and computed covariance matrices along the NEES line. The jagged appearance is due to the quantization in the sampling. $\V=\V^a$ and $\W=\W^a$ at point 149.}
\label{fig:cov_comparison0}
\end{figure}

Suppose the actual (groundtruth) process and observation noise intensities are $\V^a=1$ and $\W^a=0.1$. However, these values are not known, and a black-box tuning algorithm will try candidate values for $\V$ and $\W$. In the appendix, we derive the expressions to compute $\Jnees(\V,\W,\V^a,\W^a)$. Fig.~\ref{fig:scan_results} plots these values for different choices of $(\V,\W)$. When $\V<\V^a$ and $\W<\W^a$ (bottom left), $\Jnees$  is high because the filter is inconsistent. When $\V>\V^a$ and $\W>\W^a$ (top right), $\Jnees$ is large again because the filter is conservative. The thick curved blue line shows where $\Jnees\approx0$ and shows multiple solutions which appear consistent. The yellow curve is the set of samples of $(\V,\W)$ for which $\Jnees(\V,\W,\V^a,\W^a)\in[-0.0025,0.0025]$ ($\nees{k}(\V,\W,\V^a,\W^a)\in[1.995,2.005]$). We refer to this curve as the ``NEES line.'' Fig.~\ref{fig:cov_comparison0} plots the log determinants of $\covCond{\kLst}{\kLst}(\V,\W)$ and $\covCondTrue{\kCur}{\kLst}(\V,\W,\V^a,\W^a)$ along this curve. These results largely support Oshman and Shaviv~\cite{oshman2000optimal}: there are multiple solutions which appear to be consistent with the NEES, and the optimal solution occurs near where the covariance is minimized. However, we see it is possible to choose values which are \emph{slightly} inconsistent.

There are two implications for these results. The first is that, to compute the optimal solution, we had to derive closed form solutions for the NEES. This is possible in linear systems only by knowing the groundtruth noises, which are not available in practice, and for nonlinear systems is generally unachievable in closed form. Therefore, empirical techniques will have to be used. Second, tuning to incorrect noises means that the filter is not robust to changes in the configuration. For example, if the correct values for $\V$ and $\W$ are used, the filter should be consistent given any timestep length. Slight errors in these values no longer means this is true.

For example, consider the filter solution when $\V=1.045,\W=0.95$ which is around point 130 on Fig.~\ref{fig:cov_comparison0}. For $\Delta{t}=0.1$ this gives a the value $\Jnees=0.0018$ ($\nees{k}=2.0037$). Furthermore, if one computes the values of $\nees{\kCur}(\V,\W,\V^a,\W^a)$ using fixed values for noise intensities but varying $\Delta{t}$, there is a clear and significant change in the NEES for these various timestep lengths.

\begin{figure}
\begin{center}
\includegraphics[width=0.8\linewidth]{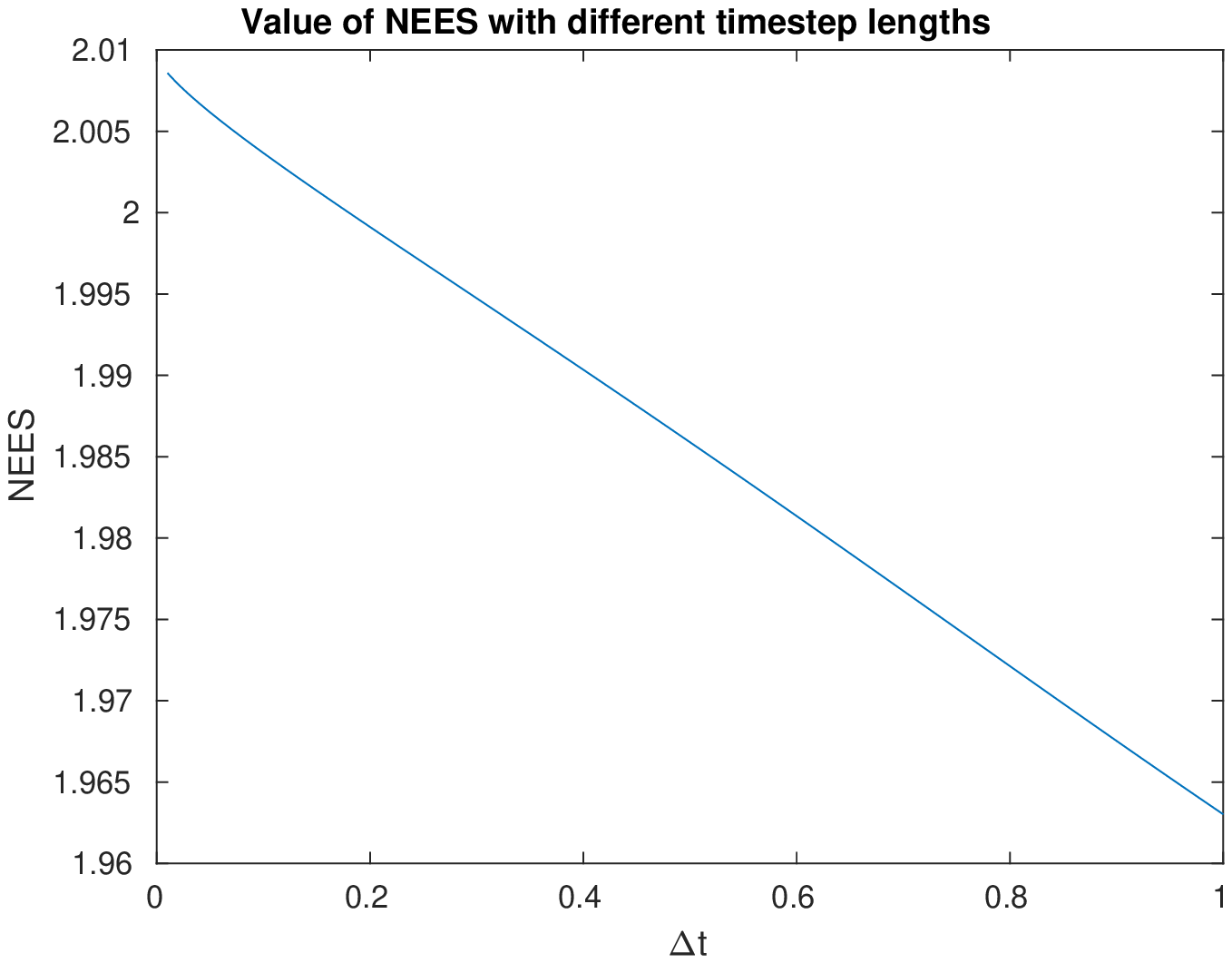}
\end{center}
\caption{$\nees{k}$ for different values of $\Delta{t}$. }
\label{fig:cov_comparison1}
\end{figure}

Our motivation is to find a way to expose the errors more clearly, since they can lead to suboptimal solutions in auto-tuning techniques. In Fig.~\ref{fig:cov_comparison1}, we compute $\nees{\kCur}(\V,\W,\V^a,\W^a)$ using fixed values for the noise intensities but varying $\Delta{t}$ between \SI{0.1}{\second} and \SI{1}{\second}. As can be seen, these results suggest that the impact of a tuning error becomes more significant if the filter timestep changes relative to the timestep used when tuning the original filter.

%As can be seen, there is a signficant change in the NEES for these timestep lengths. 

\section{Noise Tuning}

\subsection{The Effects of Noise Perturbations}

The previous section demonstrated that the $\Jnees$ values are ambiguous in supporting correct noise tuning. When coupled with minimising the covariance, the values can be found in theory; however, the differences can be small. The differences become apparent at long prediction intervals, which is computationally costly, and worse converges very slowly over lengthening intervals. However, this can suggest that one strategy is to use different timestep lengths and observe the effect on estimation statistics.

\begin{figure} [ht!]

\centering
%\subfloat[]{
\includegraphics[width=0.8\linewidth]{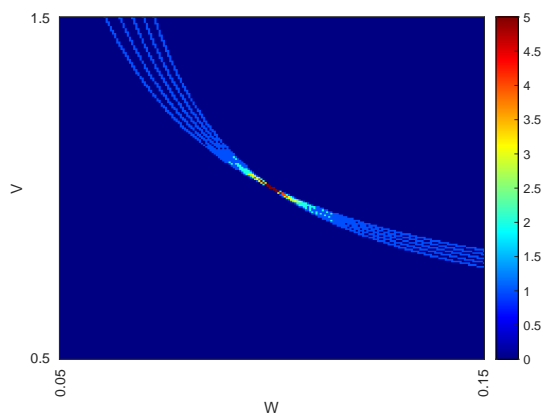}
%}
%\subfloat[]{
%\includegraphics[width=0.22\textwidth]{figs/suboptimal_tuning_nees_dR.e%ps}
%\label{fig:perturb_dR}
%}

\caption{Overlay of NEES curves with values of $\Delta{t}=[0.1,0.2,0.3,0.4,0.5]$.
%and (b) $\Delta{R}=[0,0.1,0.2,0.3,0.4]$. 
Each cell contains the count of the number of runs within which a NEES value of 2 is obtained.}
\label{fig:perturb_dT}
%\label{fig:perturbations}
\end{figure}

To motivate this, Fig.~\ref{fig:perturb_dT} shows the effect of computing over several different timesteps. For each timestep, van Loan's technique was used to construct the system and the NEES was calculated. As before, only the values close to 2 were kept. For each value of $\Delta{t}$ a different NEES curve is generated. All of the curves intersect at the same point which is the groundtruth value of the intensity. This is hardly surprising. If the filter is tuned to the groundtruth values, it should generate the same NEES irrespective of the timestep length. However, it also suggests that the observability of the optimal tuning parameters can be influenced by timestep length.

%To check this, Fig.~\ref{fig:perturb_dR} shows the effect of varying $W$. This could be simulated by taking the observations and adding noise to them. The results clearly show that, once again, the noise perturbation results in different curves, but these curves intersect with one another.

The foregoing has been conducted purely using a theoretical analysis of NEES calculations. To test the effect of this, we used 200 Monte Carlo runs and computed $\Jnees$ using \eqref{eqn:jnees}. Figs.\ \ref{fig:tracking1d_gt_dt01_log} and \ref{fig:tracking1d_gt_dt05_log} plot the $\Jnees$ values for $\Delta{t}=0.1$ and $\Delta{t}=0.5$ respectively. These show that, despite sampling noise, we see a very similar behaviour again with the curve being shifted and values along a ridge being very similar.

\subsection{Cost Function for Optimization}
The conclusion of the foregoing argument is that there is implicit dependence of $\Jnees$ as a function of $\Delta{t}$. To our knowledge, this is not very well-explored in the literature. In auto-tuning Kalman filter algorithms, the $\Jnees$ is typically evaluated conditioned upon a single value of $\Delta{t}$. Of course, the alternative, where $\Delta{t}$ is allowed to vary as a parameter to $\Jnees$, results in a computationally expensive parameter search. Yet the extreme value and implicit function theorems imply that such a minimum exists somewhere between $\Delta{t} = (0,h)$ where $h$ is ``small,'' as is typical for numerical integration and required for local truncation error to be acceptably low, and as long as there are no discontinuities in $\mathbf{F}$ or $\mathbf{S}$.

\begin{figure*} [ht!]
\centering
% \subfloat[]{
% \includegraphics[width=0.3\textwidth]{figs/dt01_surface_jnees_freesan.eps}
% %dt01_surface_jnees.eps}
% \label{fig:tracking1d_gt_dt01}
% }
% \subfloat[]{
% \includegraphics[width=0.3\textwidth]{figs/dt05_surface_jnees_freesan.eps}
% \label{fig:tracking1d_gt_dt05}
% }
% \subfloat[]{
% \includegraphics[width=0.3\textwidth]{figs/dt0105_surface_jnees_freesan.eps}
% \label{fig:tracking1d_gt_dt0105}
% }
% \\
\subfloat[]{
\includegraphics[width=0.40\textwidth]{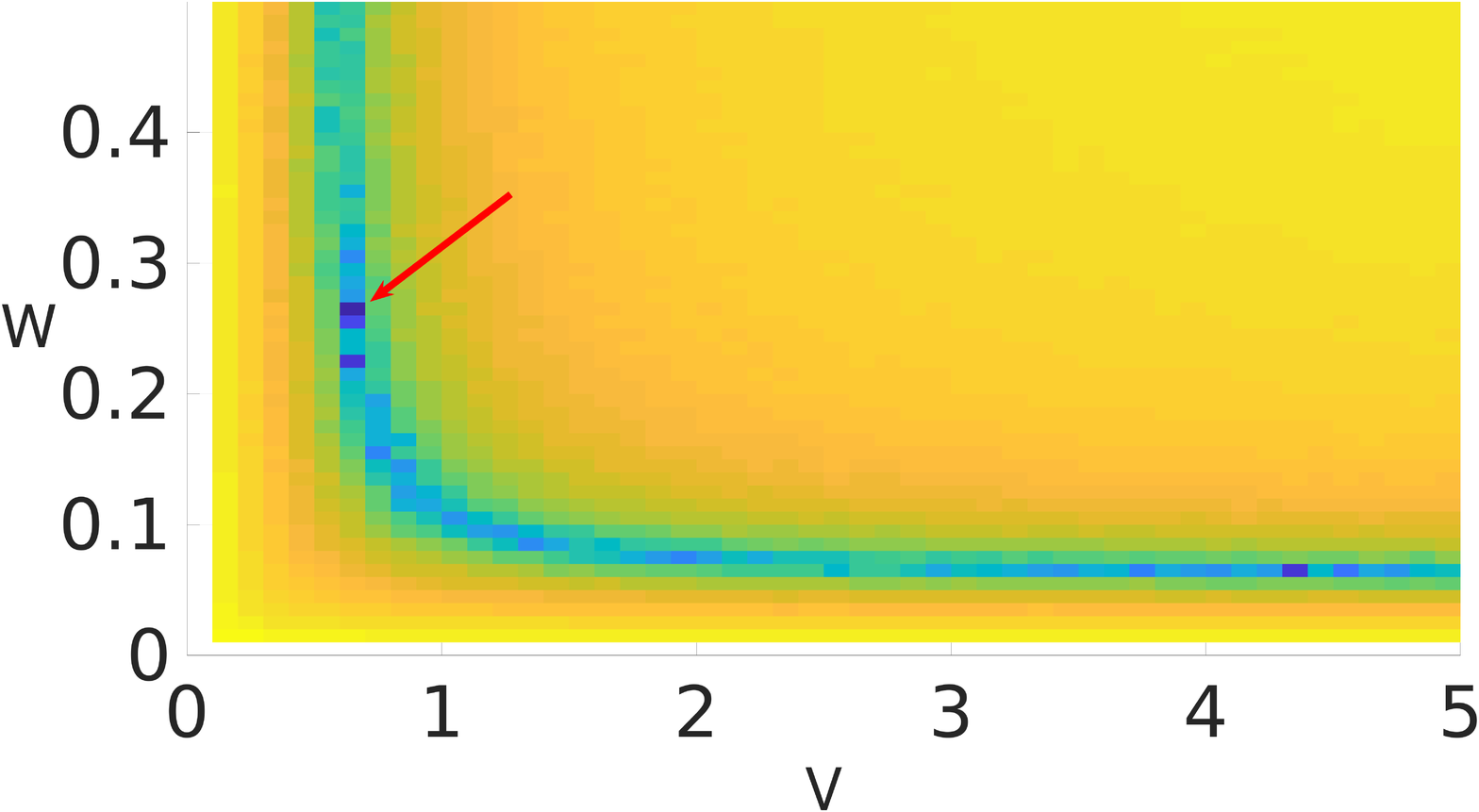}
\label{fig:tracking1d_gt_dt01_log}
}
\subfloat[]{
\includegraphics[width=0.40\textwidth]{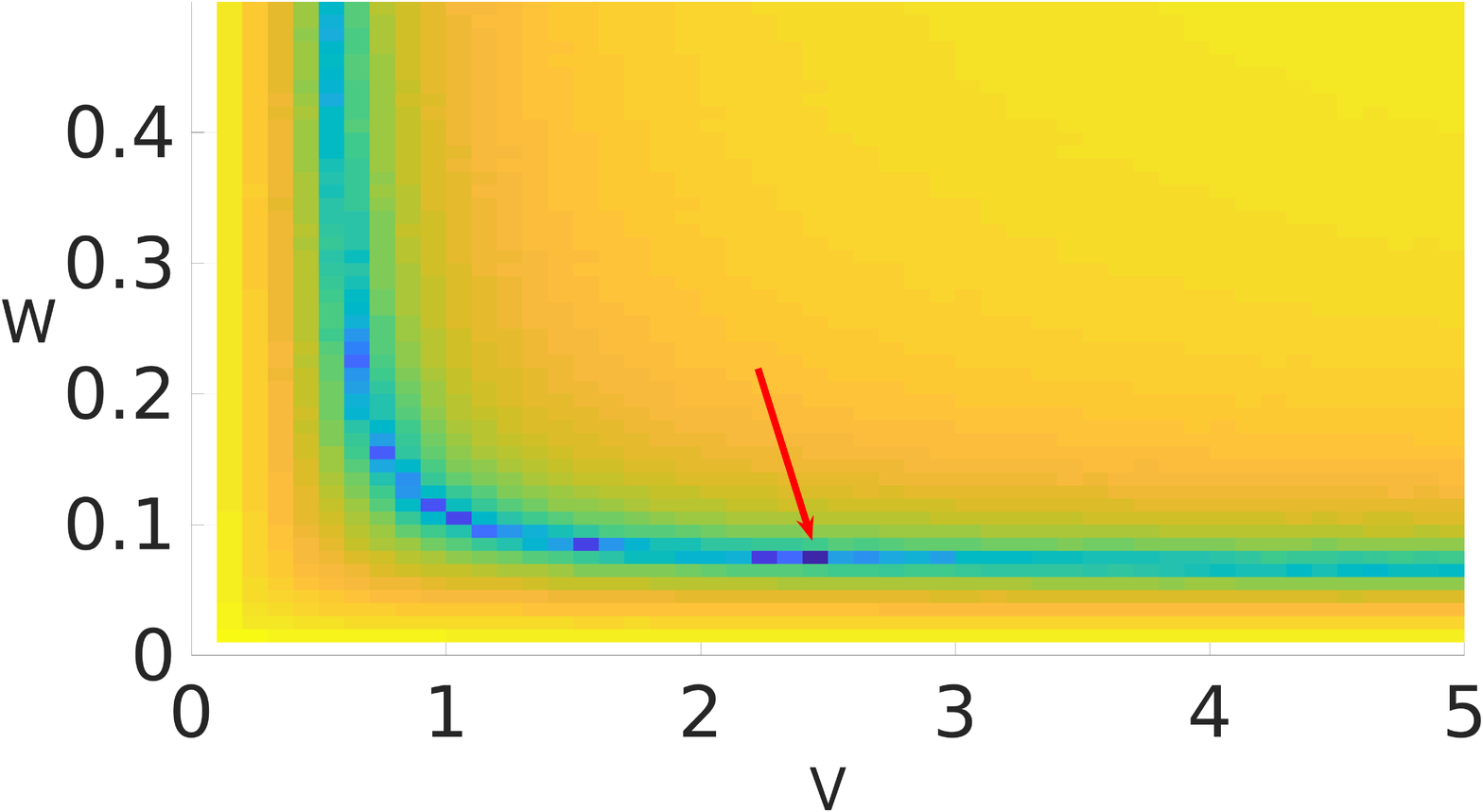}
\label{fig:tracking1d_gt_dt05_log}
}\\ 
\subfloat[]{
\includegraphics[width=0.40\textwidth]{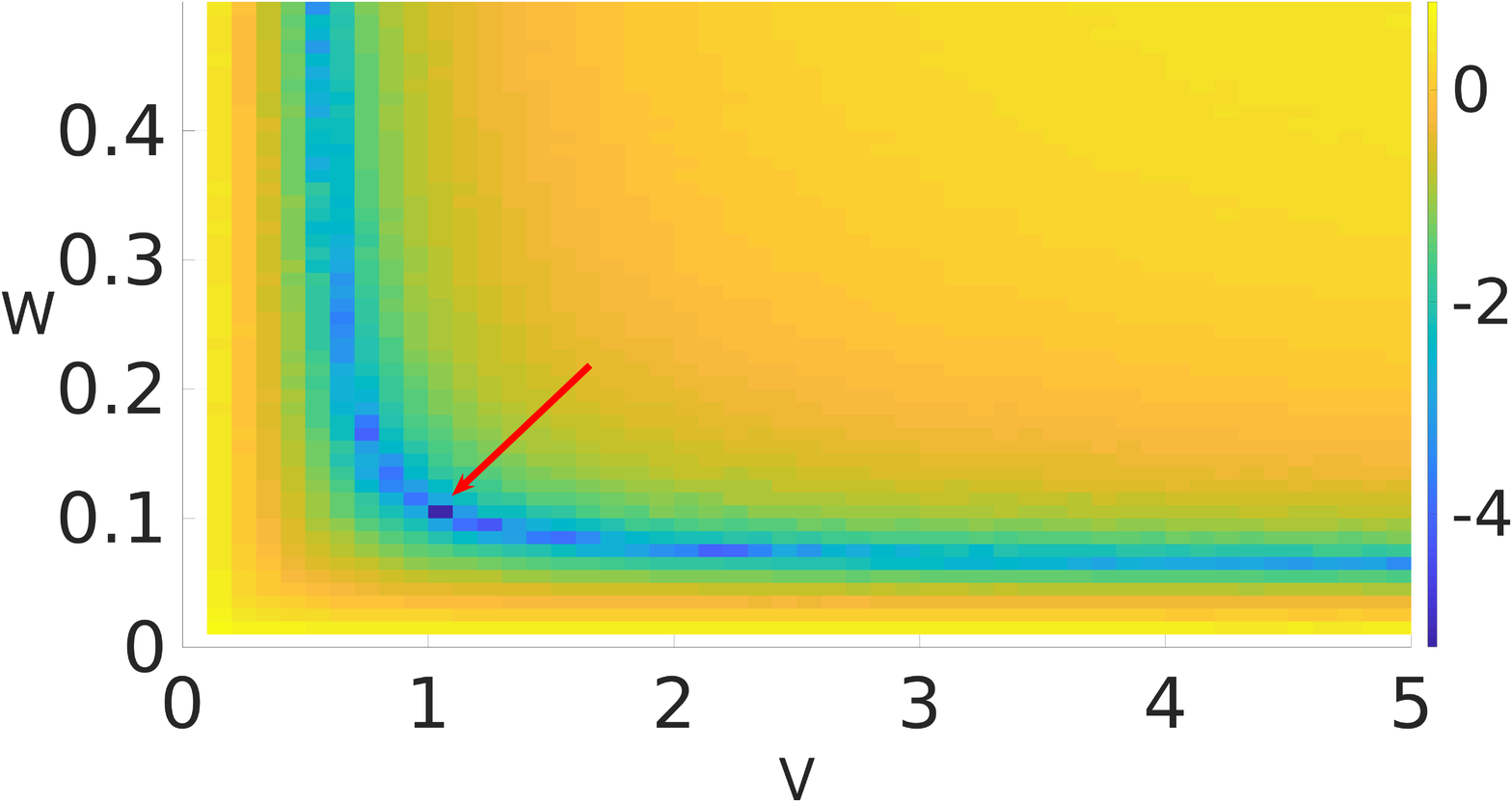}
\label{fig:tracking1d_gt_dt0105_log}
}
\caption{In Figures (a), (b), (c), the colorbar shows the value of $lg(\Jnees)$. In (a), $\Delta t = 0.1$; In (b), $\Delta t = 0.5$; In (c), we calculate $\Jnees$ at both $\Delta t$ for each $[\V,\W]$ pair but only pick the larger cost. The red arrow indicates where the $\Jnees$ is the smallest in each plot. We can see that only in (c) we have a global minima around the groundtruth value $[\V=1,\W = 0.1]$}
%In Figures (a), (b), (c), the colorbar shows the value of $\Jnees$. Figures (d), (e), (f) are the $lg(\Jnees)$ value of (a), (b), (c) respectively.  In (a), $\Delta t = 0.1$; In (b), $\Delta t = 0.5$; In (c), we calculate $\Jnees$ at both $\Delta t$ for each $[\V,\W]$ pair but only pick the larger cost. The red arrow indicates where the $\Jnees$ is the smallest in each plot. We can see that only in (c)/(f) we have a global minima around the groundtruth value $[\V=1,\W = 0.1]$} 
\label{fig:tracking1d_gt_dt_compare}
\end{figure*}

\begin{figure*} [ht!]
\centering
\subfloat[]{
\includegraphics[width=0.40\textwidth]{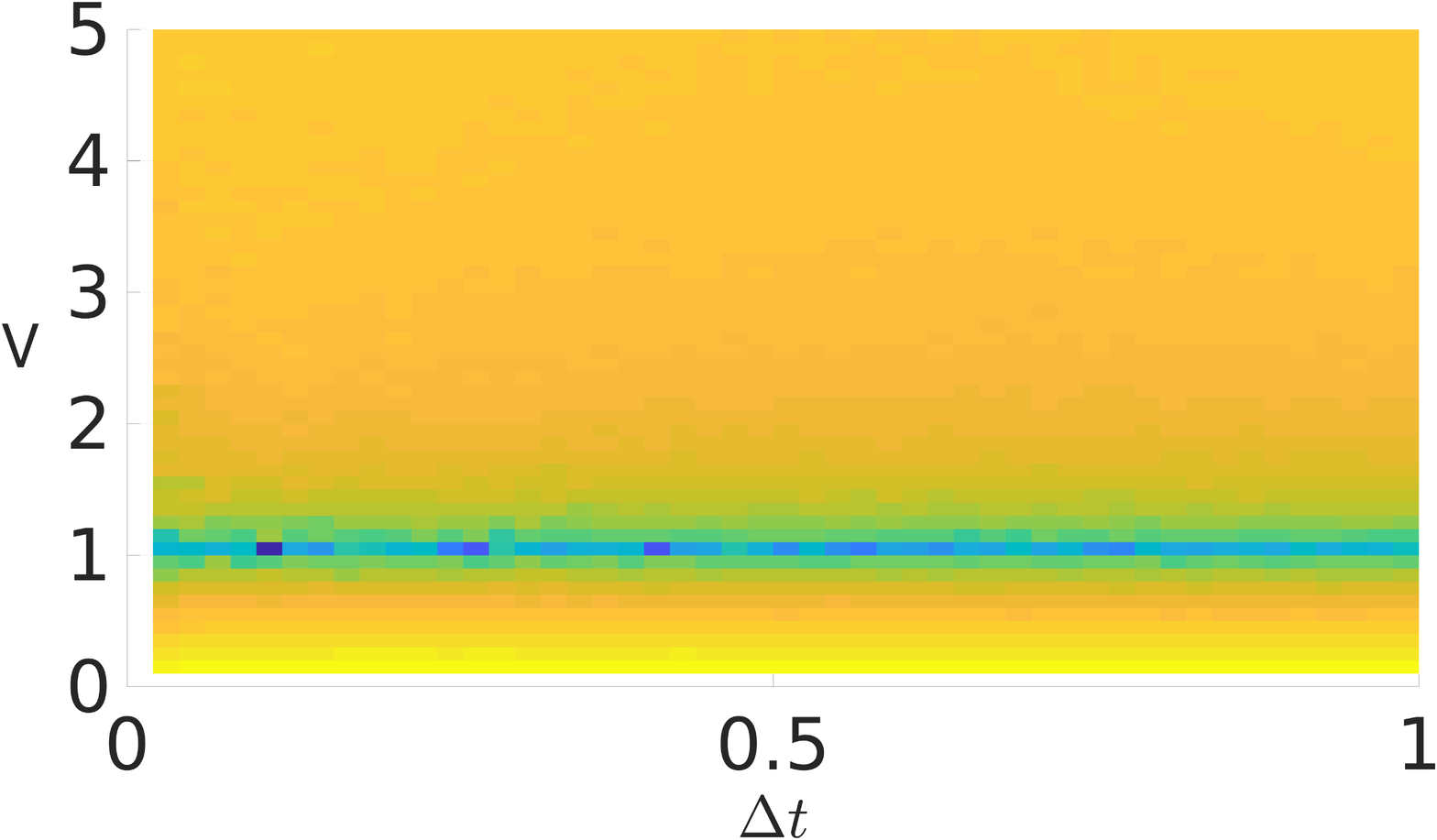}
\label{fig:tracking1d_gt_a}
}
\subfloat[]{
\includegraphics[width=0.40\textwidth]{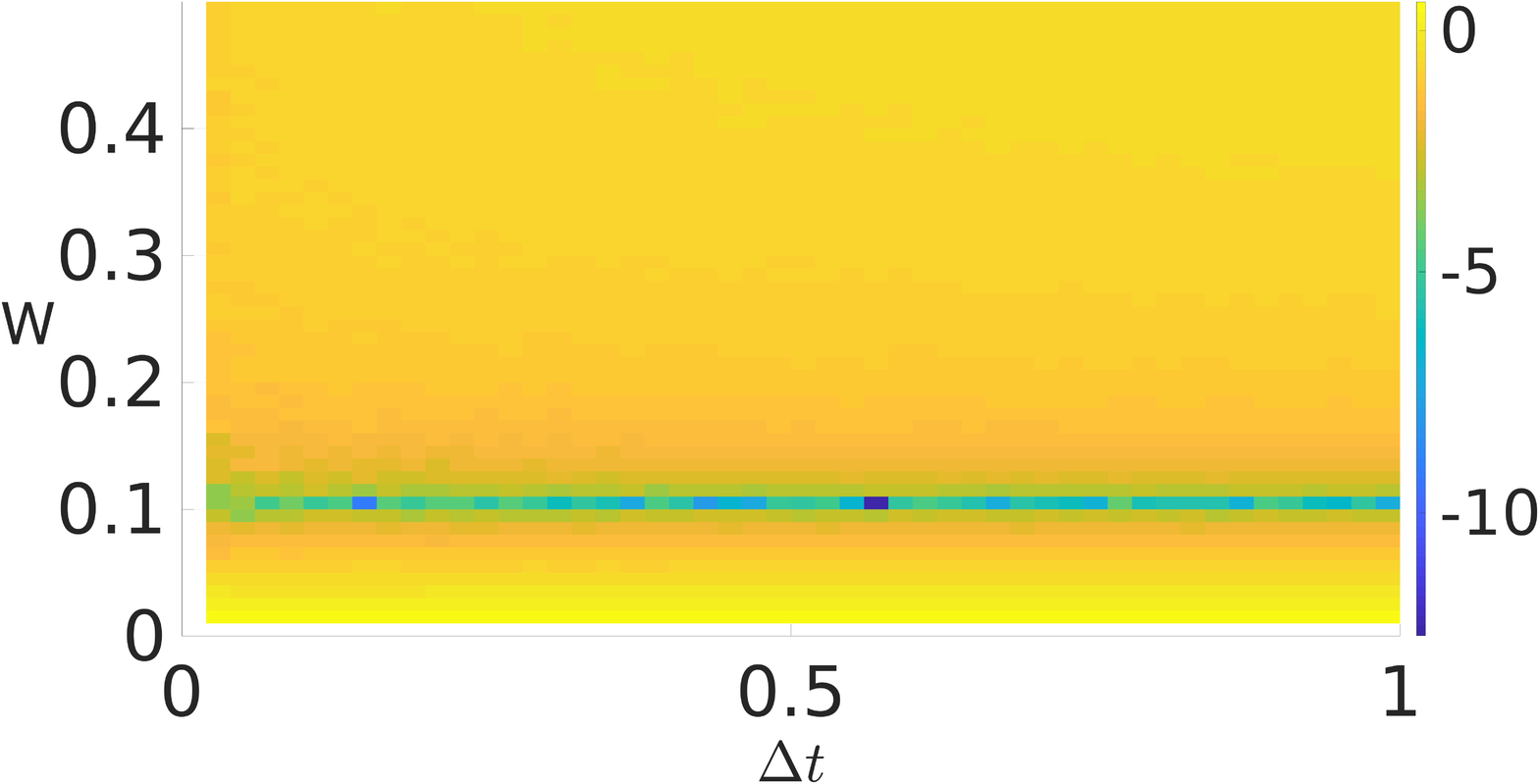}
\label{fig:tracking1d_gt_b}
}

\caption{Figure (a) fixes $\W = 0.1$ and plots $\V, \Delta t$ versus $lg(\Jnees)$. Figure (b) fixes $\V= 1$ and plots $\W, \Delta t$ versus $lg(\Jnees)$. We can see only when both $\V, \W$ are around the groundtruth value, $\Jnees$ is small whatever the $\Delta t$ is.} 
\label{fig:tracking1d_gt}
\end{figure*}

To avoid the need for an expensive search, we choose a sample of $\Delta{t}$ values and a logical operation in our search: for each pair $[\V,\W]$, $\W \in [0.01, 0.5]$, $\V \in [0.1, 5.0]$, groundtruth $\V = 1,\W = 0.1$ we calculate $\Jnees$ using $\Delta t = 0.1$ and $\Delta t = 0.5$. Then, we only record the larger $\Jnees$ and get another plot. The results are shown in Fig. \ref{fig:tracking1d_gt_dt01_log},\ref{fig:tracking1d_gt_dt05_log},\ref{fig:tracking1d_gt_dt0105_log}. Note the plots show $lg(\Jnees)$ because, in this way, $\Jnees$ smaller than 1 will be negative, its color is more clear. 
%In Fig. \ref{fig:tracking1d_gt_dt01}, the $\Jnees$ looks similar. To make it clear, we again calculate $lg(\Jnees)$ (In this way, $\Jnees$ smaller than 1 and closer to 0 will be negative so that its color is more clear).
In Fig.\ref{fig:tracking1d_gt_dt01_log}, there is a  blue curve shows the small $\Jnees$. The red arrow points out the minimum value, which is not around the groundtruth. in Figure \ref{fig:tracking1d_gt_dt05_log}, the minimum is also not at the groundtruth. We find that the global minimum $\Jnees$ is quite random when $\Delta t = 0.1$ or $\Delta t = 0.5$ or other single $\Delta t$. Thus, when we use an optimization algorithm to search the surface, the possible estimations can be quite random. However, this situation is different in case Figure \ref{fig:tracking1d_gt_dt0105_log}. The global minimum is always around [0.1,1]. It is obvious now the $\Delta t$ influences the cost function distribution. It would be interesting to see the mapping between different $\Delta t$ value and $\Jnees$, which is shown in Figure \ref{fig:tracking1d_gt}. It shows that when both $\V, \W$ are around the groundtruth value, $\Jnees$ is small whatever the $\Delta t$ is. These experiments motivates us to tune the KF with different $dt$ and find the solution that can give consistent $\Jnees$. The solution should be the close to the groundtruth. In our experiment, we found that find the solution that gives consistent $\Jnees$ with only two different $dt$ are sufficient.\\
%In Figure \ref{fig:tracking1d_gt_dt01}, when $\omega$ is fixed, the $v$ doesn't influence the $\Jnees$ value. In \ref{fig:tracking1d_gt_dt05}, where $\Delta t = 0.5$, the $\omega$ has more impact on the cost function. However, the $\Jnees$ values the deep blue curve (pointed out by those arrows) are still not distinguishable. In Figure \ref{fig:tracking1d_gt_dt0105}, the smallest $\Jnees$ is around the groundtruth $v$ and $\omega$ (where the red arrow points to). It's very clear the $\Delta t$ influence the cost function distribution. It would be interesting to see the mapping between different $\Delta t$ value and $\Jnees$.\\

\section{Experiments}

\label{sct:result}

To investigate the effects of choosing multiple sample times, we apply a Bayesian optimization (BO) auto-tuning algorithm on two linear systems, namely: a 1D tracking problem and a 2D tracking problem. In both examples, the process and measurement noises parameters are optimized together. We run two examples for the following purposes. %For the 1D tracking, one process and one measurement noise parameter are jointly optimized. For the 2D tracking problem, two process and two measurement noise parameters are optimized. We run two examples for the following purposes.

\subsection{Bayesian optimization tuning}
\begin{itemize}
    \item \textbf{1D tracking}: For the 1D (particle) tracking system we introduced before, we can see the benefits of using multiple sample time during the optimization. We display the numerical optimization result and show the process of BO, from where we can see the exploration ability of the BO.
     \item \textbf{2D tracking system}: In the 2D tracking system, we are going to optimize 4D parameters. i.e. 2 process noise parameters and 2 measurement noise parameters. We perform the $\chi^2$ test to show that the filter is consistent.
    %\item \textbf{mass-spring-damping}: In the 1D tracking system, we only provide one groundtruth pair. In the mass-spring-damping system, we do optimization on multiple groundtruth pairs. We demonstrate that our system can be robust whatever the noise parameters are. Especially when the noise is large, the optimization become more challengeable.
\end{itemize}
\indent We use our previous work's optimization process \cite{chen2018weak}. i.e. GPBO (Gaussian Process BO). However, now we run the Kalman filter ($N$ Monte Carlo simulations) with two sample time ($\Delta t$ = 0.1 , $\Delta t = 0.5$) for each set of the noise estimation. We pick the larger cost and feed it into the BO. The motivation is that we want the cost remain small with different sample time.\\
\indent Results are compared from four auto-tuning strategies. The first one is the proposed GPBO algorithm with the $\Jnees$ cost function. To assess the value of the multiple sample time strategy, we compare it to our previous approach, where we use $\Delta t= 0.1$ only. To further extend our previous work, we compare the GPBO with the Downhill Simplex (DS) algorithm. From Figure \ref{fig:tracking1d_gt_dt0105_log}, we can see that even the groundtruth is at the correct position, the cost along the blue curve is close to each other, which brings a challenge to the optimizer. We show that the GPBO can efficiently explore the cost surface and achieve better results than the Downhill Simplex algorithm.
%To assess the value of the Student's-t process surrogate modeling and multi-time scale sampling approaches, we also compare to the previously developed Gaussian Process-based BO, using the $\Jnis$ cost function and only using $\Delta t = 0.1$ as the sample time. Finally, %we show our algorithm's strength with others, 
%we also compare to the results from the downhill simplex auto-tuning algorithm \cite{powell2002automated}, using $\Jnees$ as the cost function and also taking advantage of different $\Delta t$ sample times to make the comparison fair. %To make the comparison fair, the downhill simplex algorithm use $\Jnees$ as cost function and also take advantage of different $\Delta t$
After optimization convergence of each method across 200 Monte Carlo runs, the following are evaluated to compare the resulting filter tuning solutions: the numerical value of the optimized noise parameters; filter dynamic consistency, i.e. the error between the groundtruth state and the estimation should be within a threshold $\sigma$; and BO surrogate model visualizations, to demonstrate the solution search process.
\begin{figure*} [ht!]
\centering
\subfloat[]{
\includegraphics[width=0.41\textwidth]{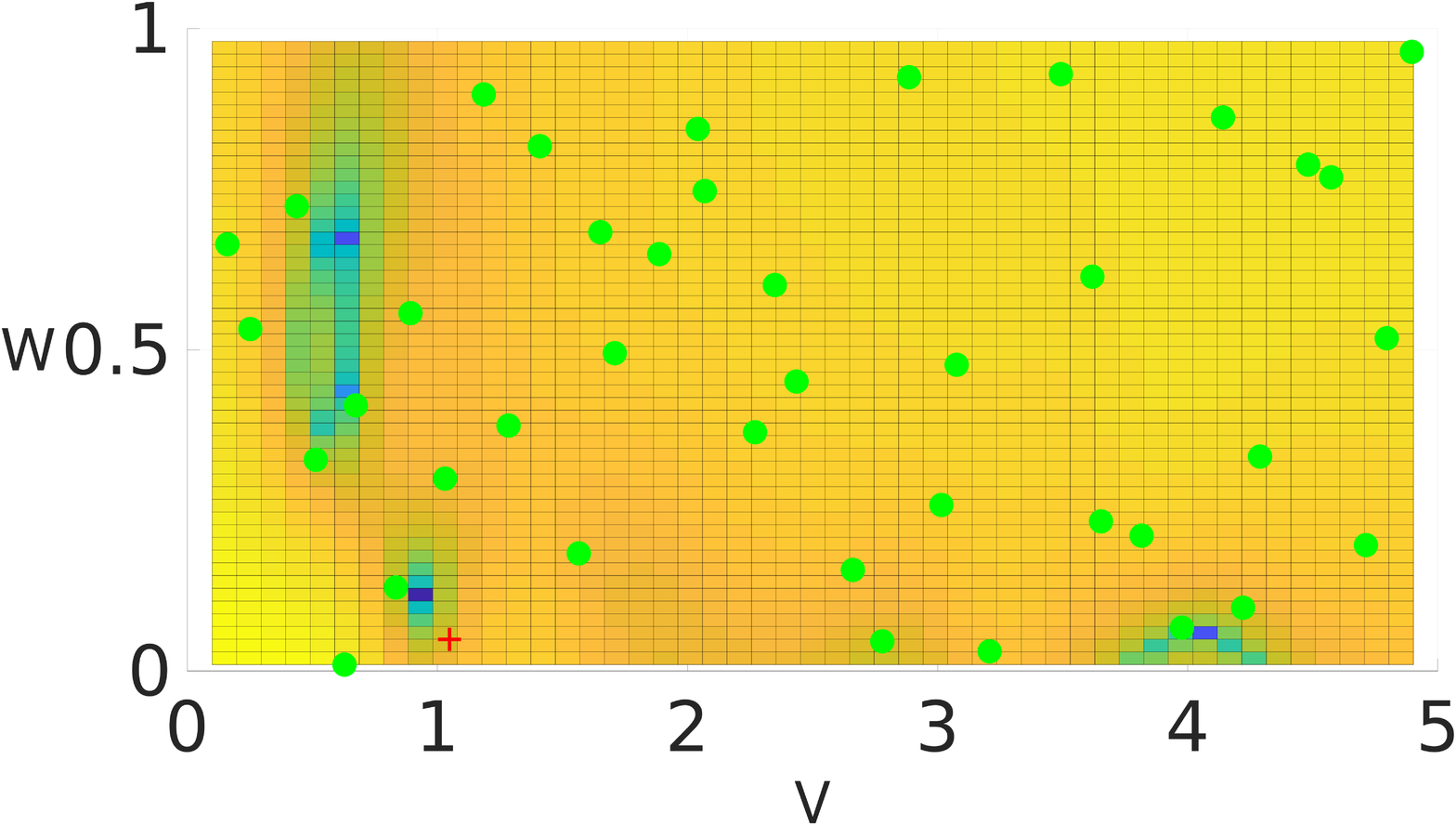}
\label{fig:tracking1d_bayesopt_a}
}
\subfloat[]{
\includegraphics[width=0.41\textwidth]{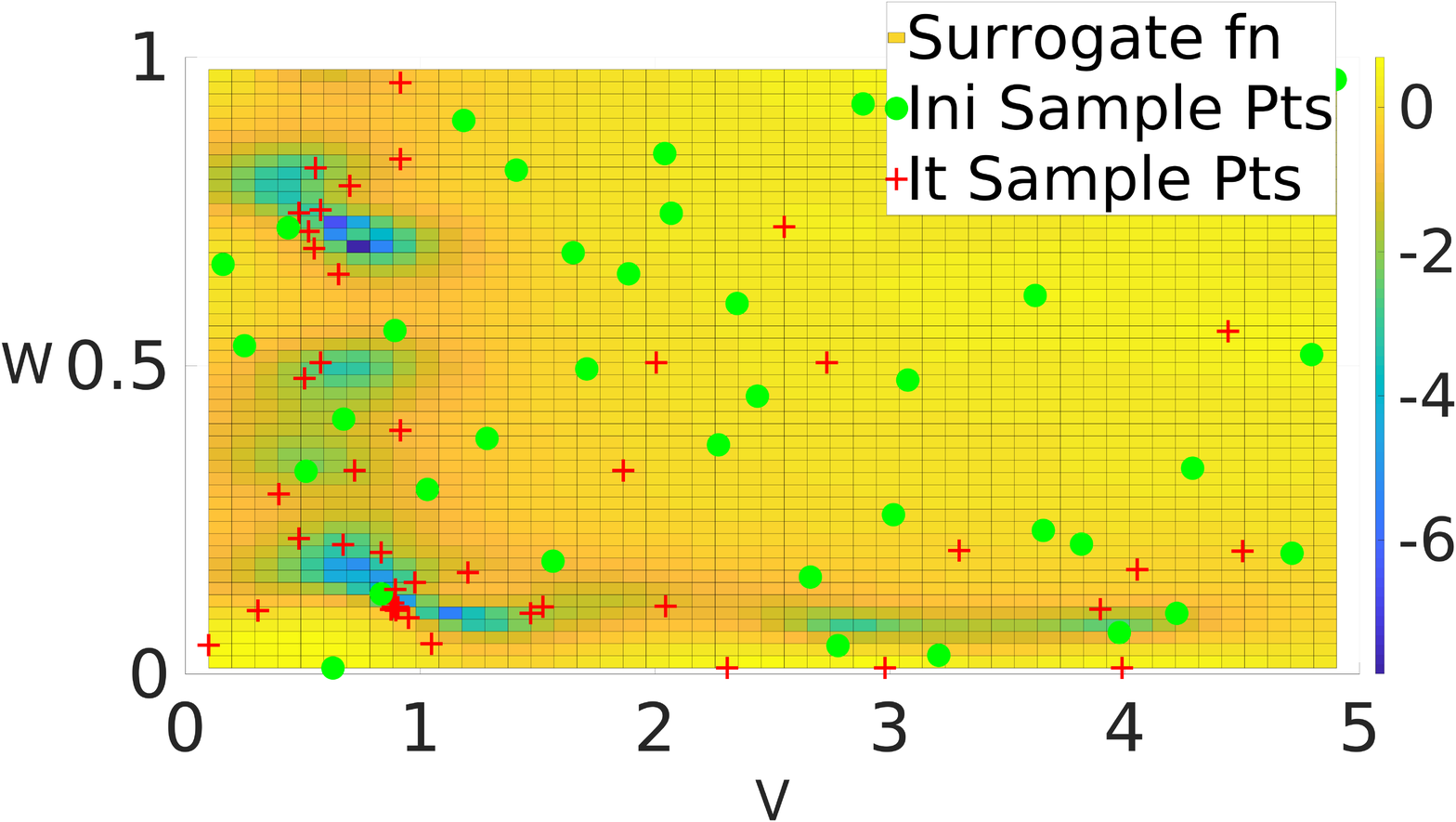}
\label{fig:tracking1d_bayesopt_b}
}\\
\subfloat[]{
\includegraphics[width=0.41\textwidth]{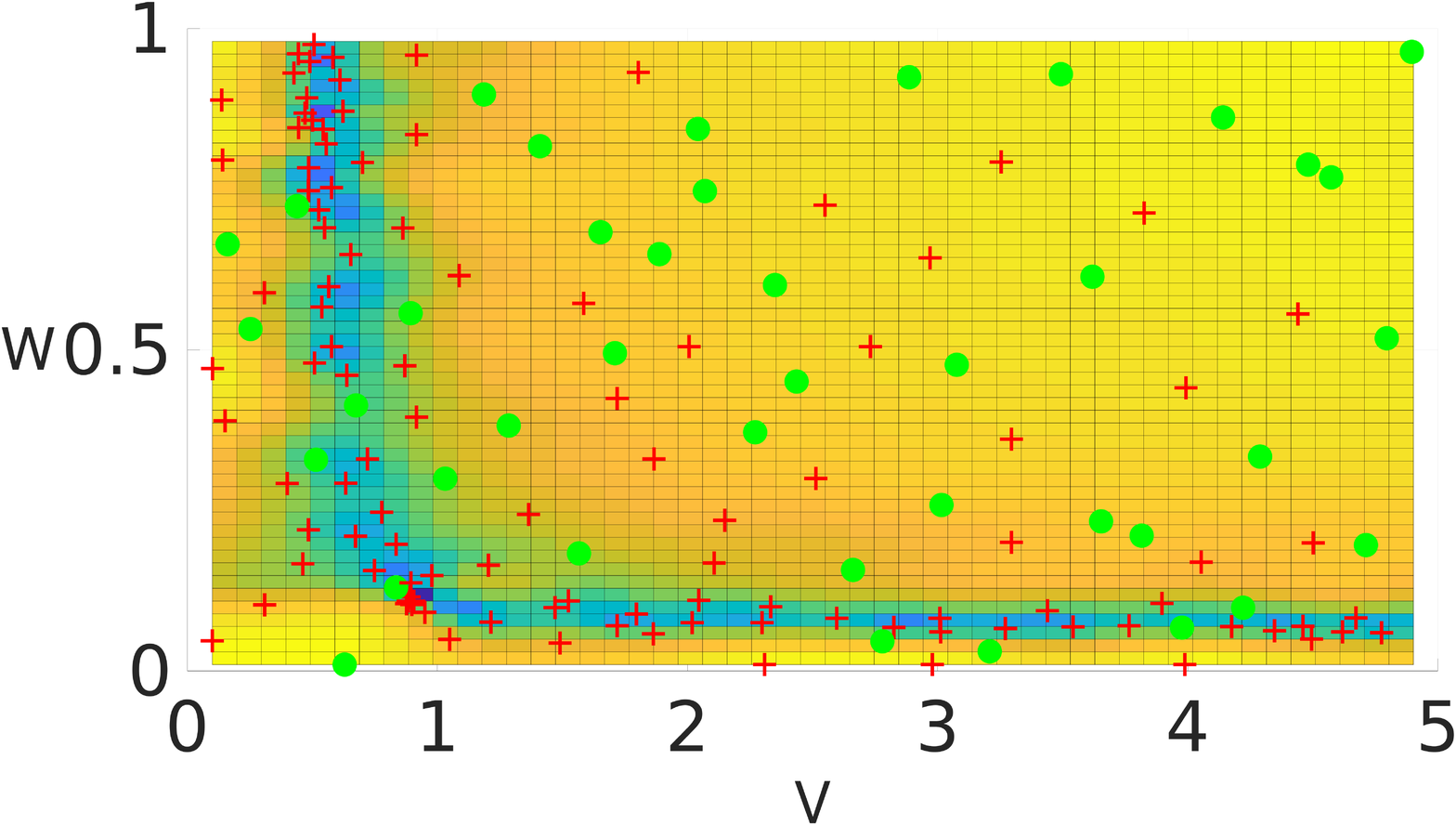}
\label{fig:tracking1d_bayesopt_c}
}
\subfloat[]{
\includegraphics[width=0.41\textwidth]{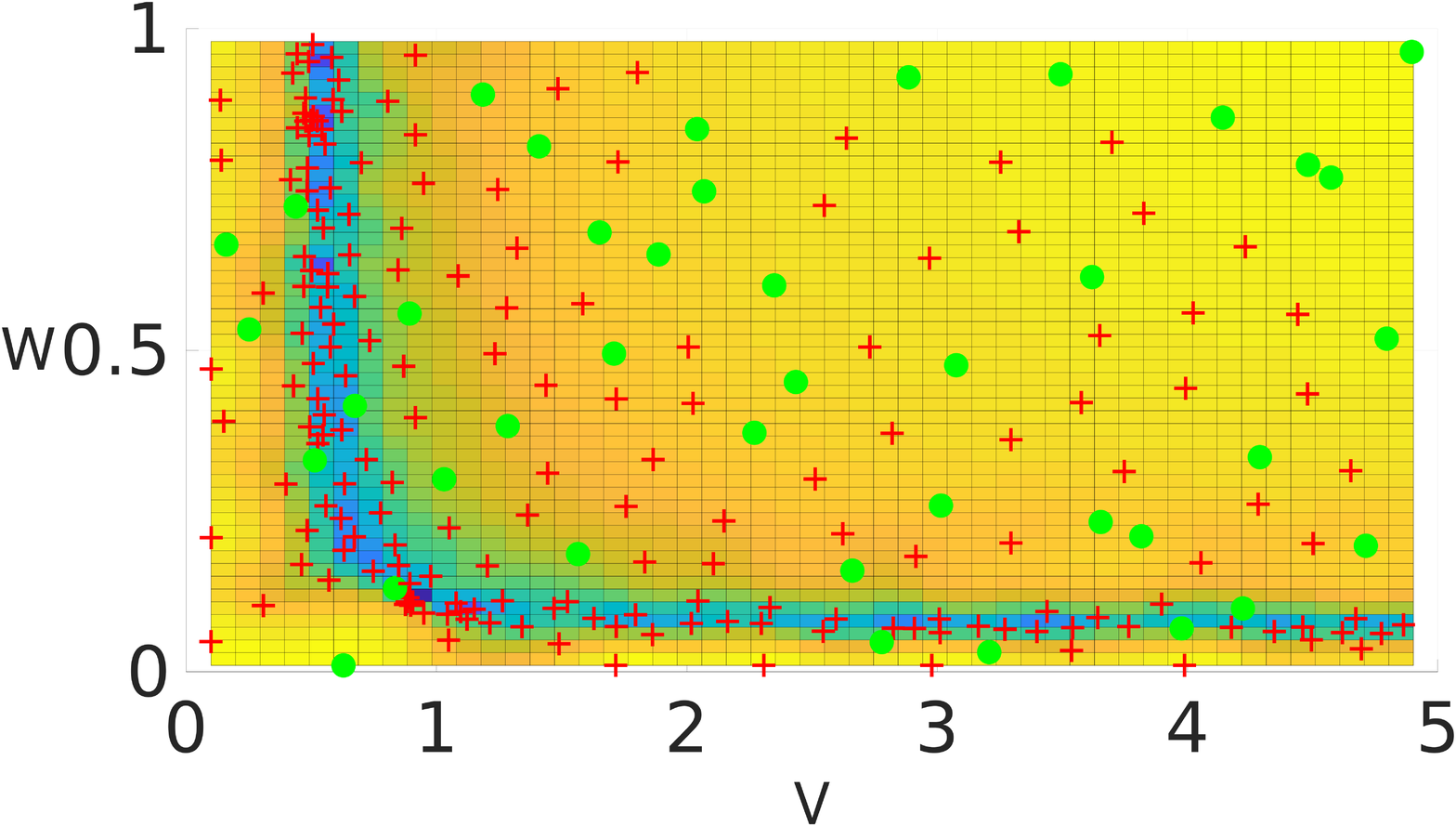}
\label{fig:tracking1d_bayesopt_d}
}
\caption{ GPBO surrogate model for $\Jnees$ cost, showing initial random sample points (green dots) and estimations (red crosses) infered by GPBO in different iterations. From (a) to (d), with more and more estimations, our algorithm successfully explore the cost space. The final surrogate model is similar to the real cost surface from Figure \ref{fig:tracking1d_gt_dt0105_log}. Finally, it finds the minimum around $v=1$ and $w=0.1$.} 
\label{fig:tracking1d_bayesopt}
\end{figure*}
\begin{table*}[!htbp]
\centering
\caption{Tracking 1D Optimization result}
\begin{tabular}{ *3c | *2c | *2c | *2c | c}
\toprule
& \multicolumn{2}{c}{GPBO, $\Delta t = 0.1, 0.5$} & \multicolumn{2}{c}{GPBO, $\Delta t = 0.1$} & \multicolumn{2}{c}{DS $\Delta t = 0.1, 0.5$} & \multicolumn{2}{c}{DS $\Delta t = 0.1$}&
\multicolumn{1}{c}{Groundtruth}\\
& $v$ & $w$ & $v$ & $w$ & $v$ & $w$ & $v$ & $w$ \\
\midrule
%Run1 & 0.805 & 0.121  & 1.833 & 0.069 &  0.575 & 0.260 & 0.332 & 0.070\\
%Run2 & 1.11 & 0.093  & 4.175 &0.059  & 0.822 & 0.081 & 0.793 & 0.195 \\
%Run3 & 0.636 & 0.203  & 0.517 & 0.429  & 0.456 & 0.289 & 0.221 & 0.205 & $v$ = 1\\
%\dots & ... & ... & ... & ... & ... & ...& ... & ... \\
Mean & \textbf{0.958} & \textbf{0.152} & 1.682 & 0.296  &0.602 &0.182  &0.317 & 0.145 & $w$ = 0.1\\
Variance & \textbf{0.115} & \textbf{0.010} &2.043 &0.076 &0.094 &0.011 &0.412 &0.012 & $v$ = 1\\
\bottomrule
\end{tabular}
\label{table:tracking1d_result}
\end{table*}
\subsection{1D tracking system}
The BO searching range for $\V$ is $[0.1,5]$ and $\W$ is $[0.01, 0.5]$. Two sampling periods ($\Delta t=0.1s$, $\Delta t=0.5s$) were used. In the real world implementation, we should choose the two sample times as different as possible. Each Monte Carlo run was carried out for $T = 200\Delta{t}$. For the kernel function, the Mat\'ern Kernel \cite{minasny2005matern} with $\nu=3$ and automatic relevance determination (ARD) was used. For remaining parameters such as the kernel mean, kernel hyperparmeter re-learn iteration number and the acquisition function optimization number, default values from the BO library \cite{JMLR:v15:martinezcantin14a} are used. %For the downhill simplex method, there are four main parameters, reflection($re$) expansion($ex$), contraction($co$) and full contraction($fc$). $re = 1, ex = 1, co = 0.5, fc = 0.5$. 
%However, without losing generality, we list the five most important parameters we used.
% \begin{itemize}
%     \item The surrogate model is Student-t model.
%     \item Kernel function \cite{Rasmussen2006}, \cite{genton2001classes}. We choose Matern Kernel \cite{minasny2005matern} with automatic relevance determination (ARD). ARD uses independent parameters for every dimension of a problem. We also assume the optimized parameters are independent. There is one more parameter $\nu$ in Matern Kernel, which is used to decide how many times the kernel function are differentiable. In our experiment, either $\nu = 3$ or $\nu = 5$ can result in a promising result.
%     \item We use expected improvement as our acquisition function.
%     \item Initial sample number $N_{seed}$. This is decided exprimentally. For 1D or 2D dimension, $N_{seed}=20$ should be enough. For $N_{d}$D, we recommend $N_{seed} = 30N_{d}$.
%     \item Final iteration number $N_{final}$. This is also decided exprimentally. We recommend $N_{final} = 50N_{d}$ at least unless the result is converged.
% \end{itemize}

%\indent Based on the previous setting, 
GPBO was performed 50 times to optimize $\V$ and $\W$. The results are shown in Table \ref{table:tracking1d_result}. 
From the table we can see our optimization appears robust: the estimation variance is small and the mean is close to the groundtruth value, which is a significant improvement from our previous GPBO method. Note also that the estimation has a large variance owing to the simulations' stochasticity. The downhill simplex algorithm, as expected, can get trapped in different local minima because we initialize the sample at different points. Even with the multiple timestep strategy, the downhill simplex struggles to converge to the groundtruth. An effective optimizer must explore different regions of parameter space to find the global minima, a strength of BO. Figure \ref{fig:tracking1d_bayesopt} shows the convergence of the resulting GPBO surrogate function and the set of sampled $v$ and $w$ parameters across 200 iterations. From Figure \ref{fig:tracking1d_bayesopt}, we can see as the number of iterations increases, GPBO explores increasingly around the local optimum. Finally, the optimal solution is found around $\V=1$, $\W=0.1$.

\begin{figure*} [ht!]
\centering
\subfloat[]{
\includegraphics[width=0.4\textwidth]{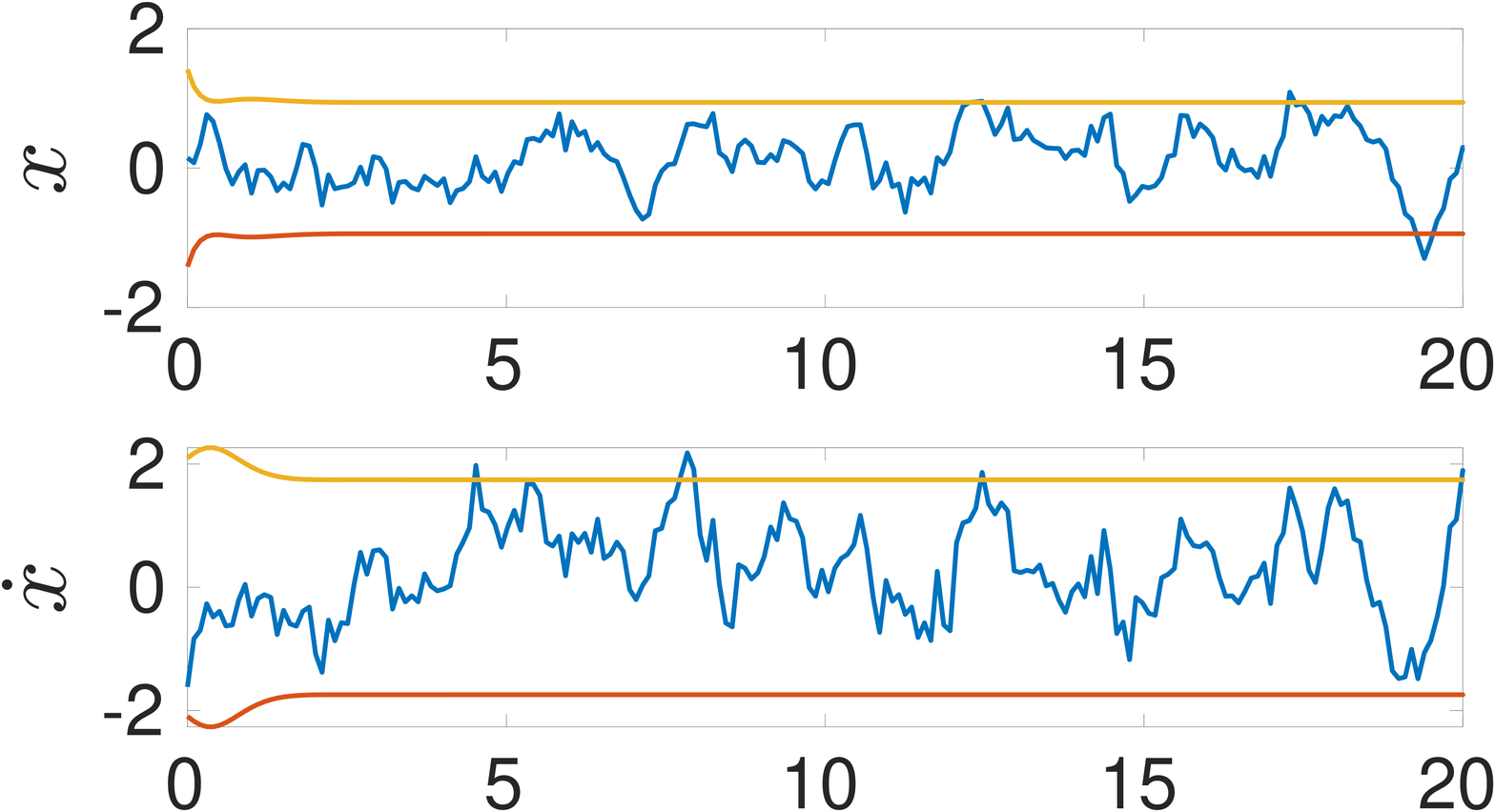}
\label{fig:con_check_tracking1d}
}
% \subfloat[]{
% \includegraphics[width=0.3\textwidth]{figs/consistency_check_msd.eps}
% \label{fig:con_check_msd}
% }
\subfloat[]{
\includegraphics[width=0.4\textwidth]{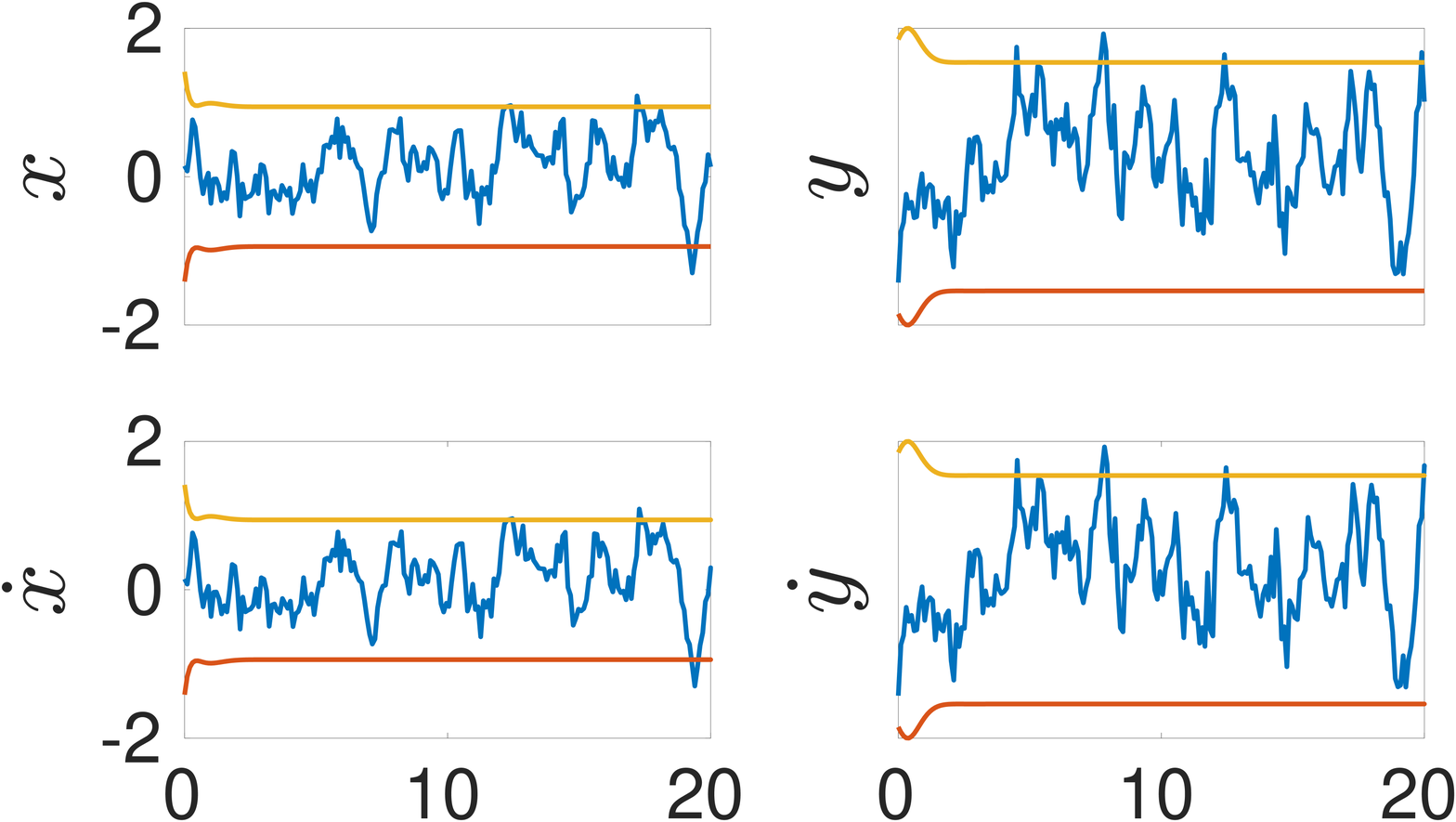}
\label{fig:con_check_tracking2d}
}
\caption{Orange lines: $2\sigma$ bounds; blue line: error between the estimated states and the real states in KF's each step. If the system is consistent, around 95$\%$ error should be within $2\sigma$ range. (a) is from the estimation result of 1D tracking system. (c) is from 2D tracking system.} 
\label{fig:consistency_check}
\end{figure*}

\begin{figure*} [ht!]
\centering
\subfloat[]{
\includegraphics[width=0.24\textwidth]{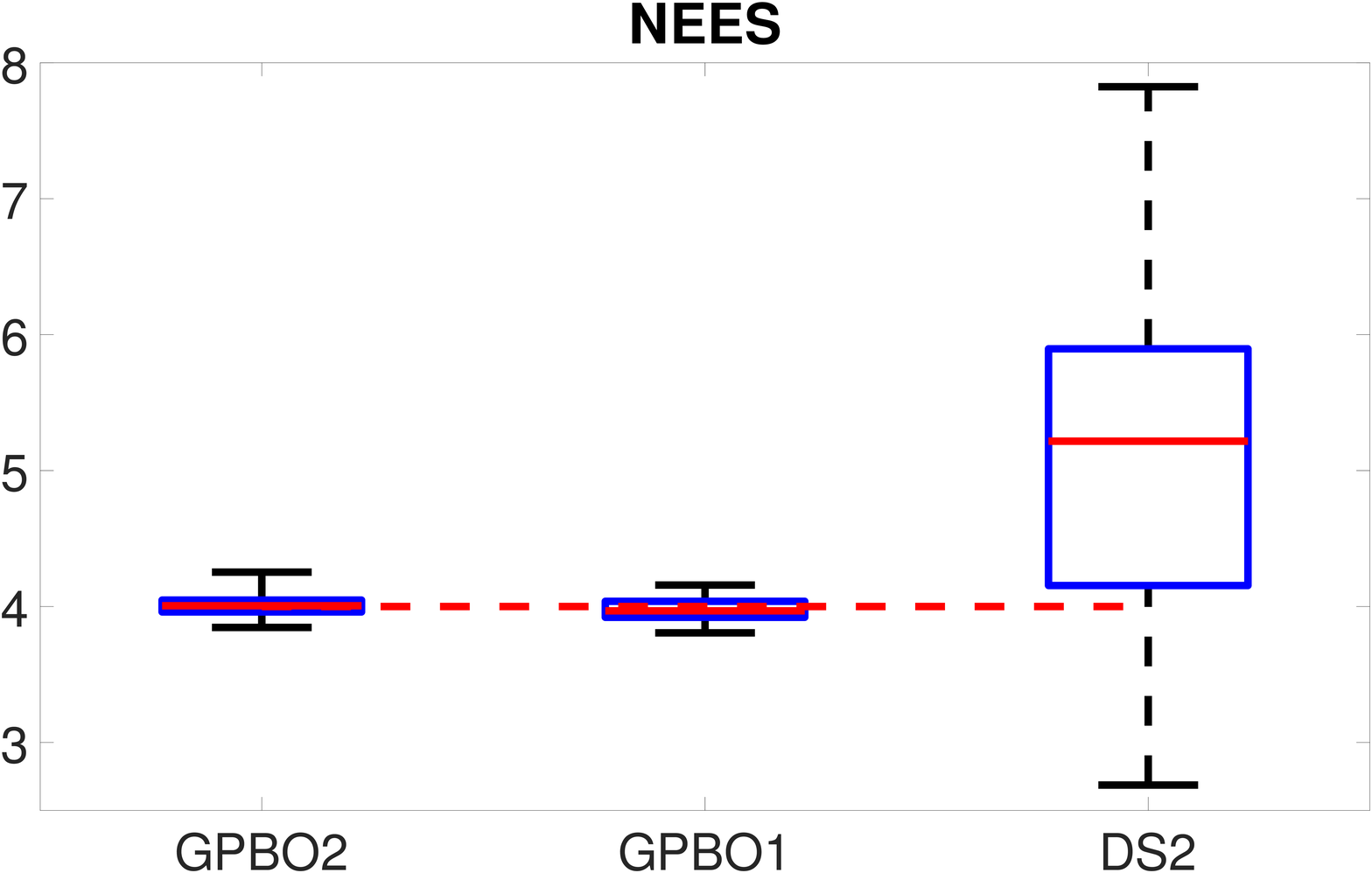}
\label{fig:nees_compare_dt01}
}
\subfloat[]{
\includegraphics[width=0.24\textwidth]{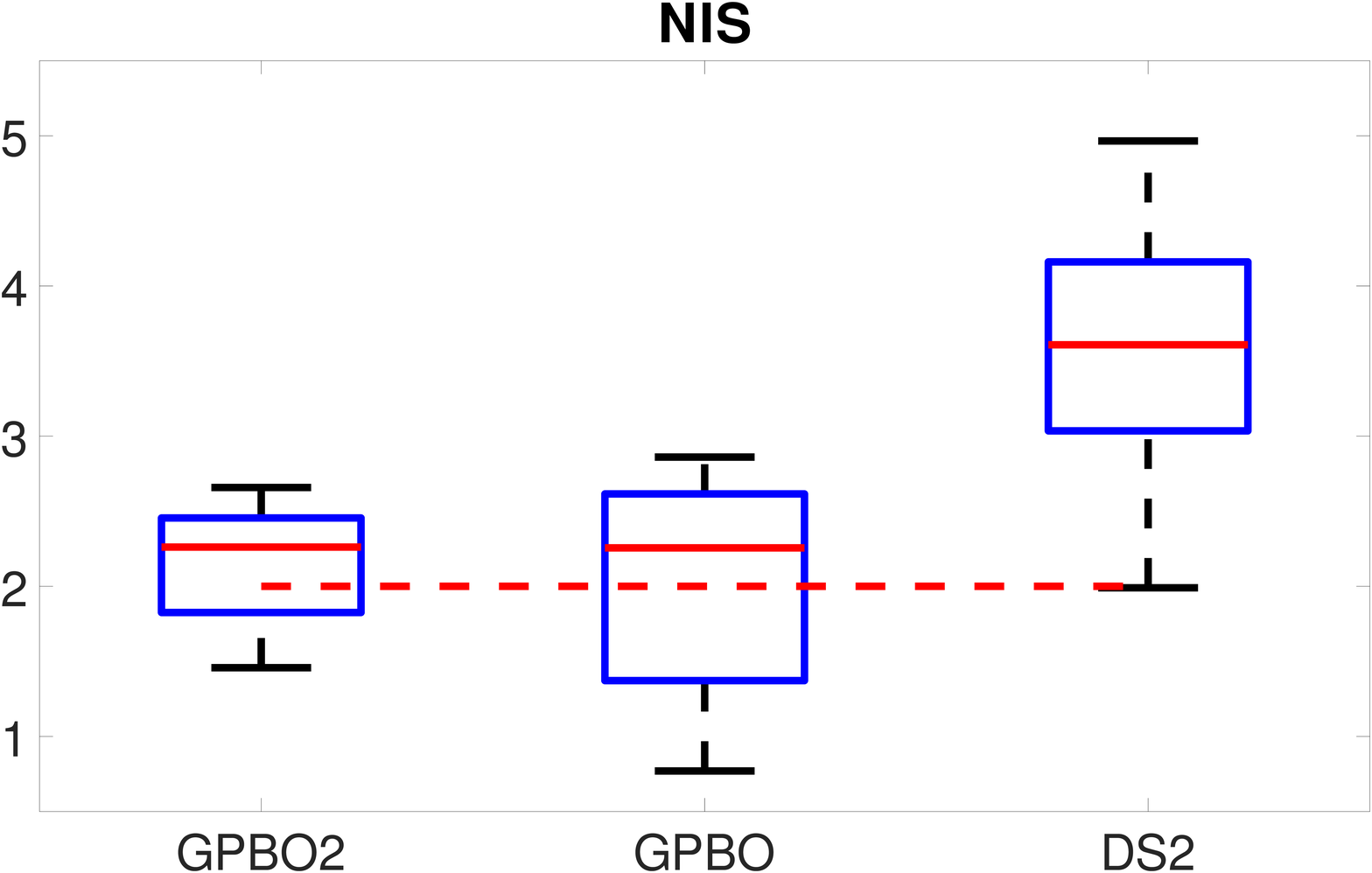}
\label{fig:nis_compare_dt01}
}
\subfloat[]{
\includegraphics[width=0.24\textwidth]{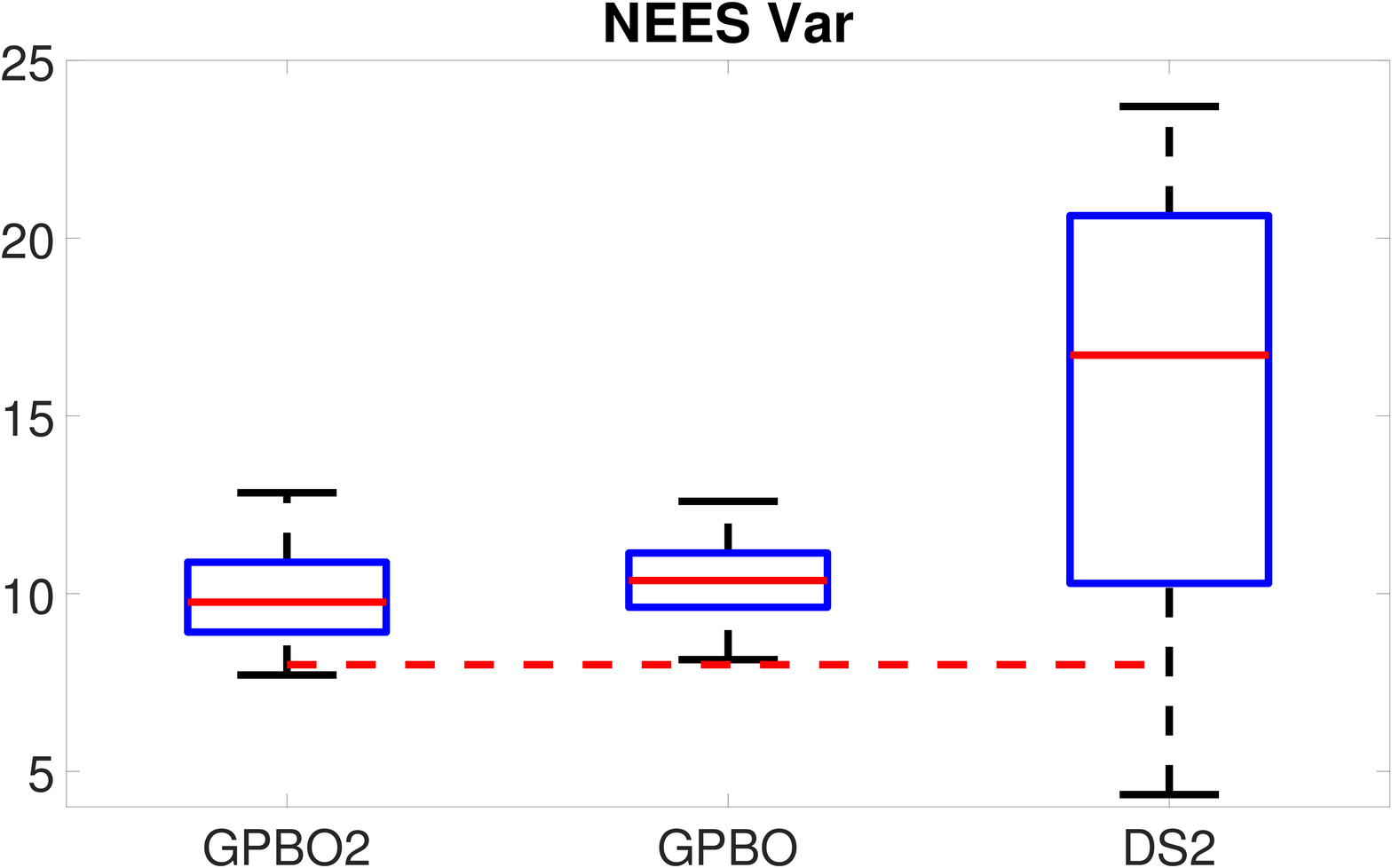}
\label{fig:nees_var_compare_dt01}
}
\subfloat[]{
\includegraphics[width=0.24\textwidth]{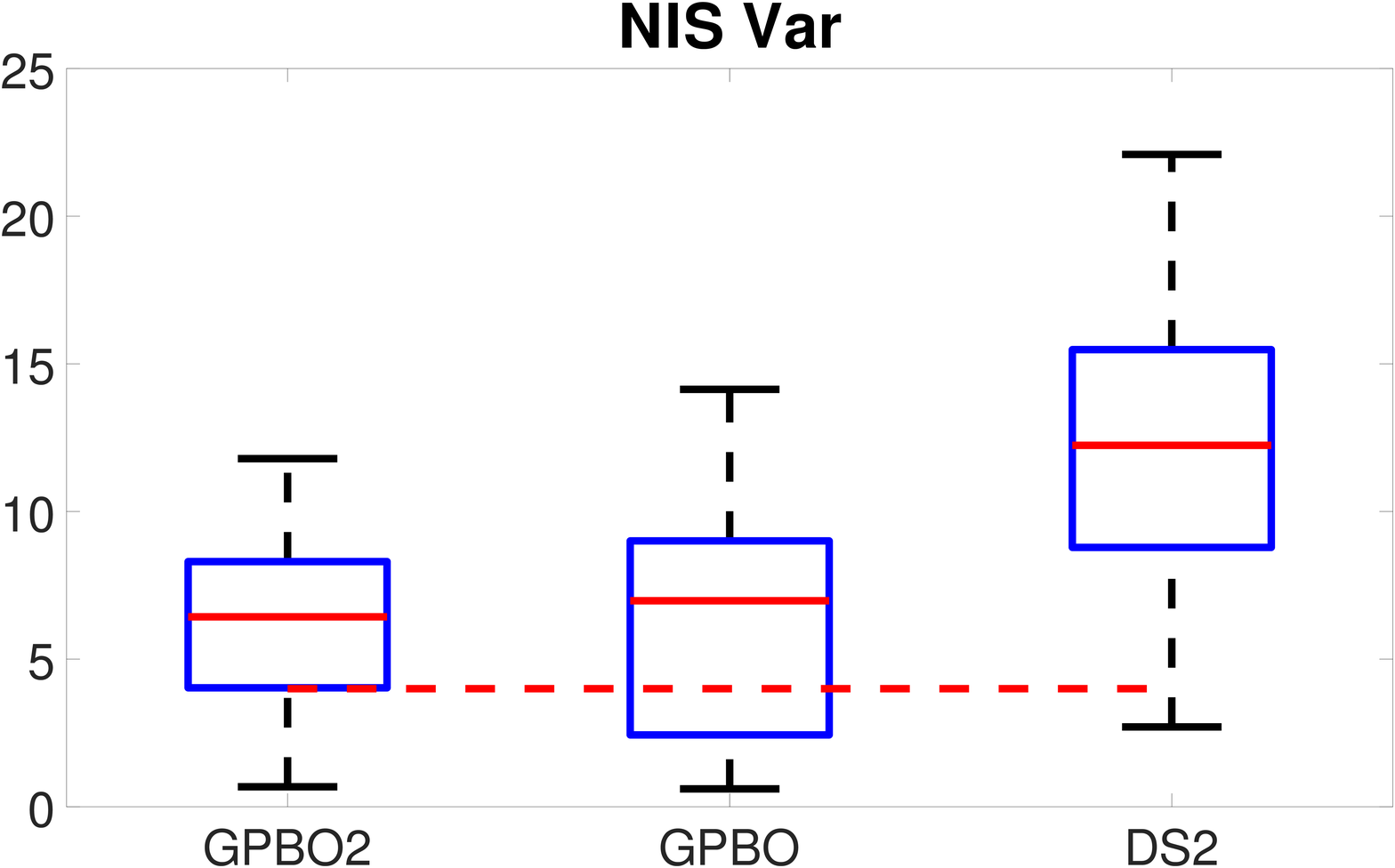}
\label{fig:nis_var_compare_dt01}
}\\
% \subfloat{
% \includegraphics[width=0.22\textwidth]{figs/nis_compare_dt04.eps}
% \label{fig:nis_compare_dt04}
% }
% \subfloat{
% \includegraphics[width=0.22\textwidth]{figs/nees_compare_dt04.eps}
% \label{fig:nees_compare_dt04}
% }
% \subfloat{
% \includegraphics[width=0.22\textwidth]{figs/nis_var_compare_dt04.eps}
% \label{fig:nis_var_compare_dt04}
% }
% \subfloat{
% \includegraphics[width=0.22\textwidth]{figs/nees_var_compare_dt04.eps}
% \label{fig:nees_var_compare_dt04}
% }\\
% \subfloat{
% \includegraphics[width=0.22\textwidth]{figs/nis_compare_dt08.eps}
% \label{fig:nis_compare_dt08}
% }
% \subfloat{
% \includegraphics[width=0.22\textwidth]{figs/nees_compare_dt08.eps}
% \label{fig:nees_compare_dt08}
% }
% \subfloat{
% \includegraphics[width=0.22\textwidth]{figs/nis_var_compare_dt08.eps}
% \label{fig:nis_var_compare_dt08}
% }
% \subfloat{
% \includegraphics[width=0.22\textwidth]{figs/nees_var_compare_dt08.eps}
% \label{fig:nees_var_compare_dt08}
% }
\caption{For each method's 50 runs result, we apply them to the Kalman filter, record the $\E{\avgnis{k}}, \E{\avgnees{k}}, \E{\avgnis{k}\avgnis{k}}, \E{\avgnees{k}\avgnees{k}}$ and plot the box plot. Red dash line: Expected value. GPBO2: GPBO with two sample time approach; GPBO1: GPBO with sample time at 0.1; DS2: Downhill Simplex with two sample time approach. %If the filter is statistically consistent, the constraints for the Eq. \ref{eqn:nees_nis} and Eq. \ref{eqn:nees_nis_cov} should be satisfied. The red line is the expected value. 
%Each row's data are collected at $dt = 0.1, dt = 0.4, dt = 0.8$ respectively. 
%If we just check the NEES value, the TPBO with $\Jnees$ method is not the best. However, combing all the plots of $\E{\avgnis{k}}, \E{\avgnees{k}}, \E{\avgnis{k}\avgnis{k}}, \E{\avgnees{k}\avgnees{k}}$ and different sample time, we can notice that the result from the TPBO with $\Jnees$ is the most robust to maintain the $\chi^2$ distribution constraints of both NIS and NEES value.
} 
\label{fig:nis_nees_chi_square_test}
\end{figure*}
\subsection{2D tracking system}
So far, we have only considered the motion of a 1D particle which required two scalar intensity values. However, our method directly extends to vector-valued intensity values. There, in this section we demonstrate the performance of the approach in a 2D tracking system, where the state is $\x{} = [x, y, \dot{x}, \dot{y}]^T$. We assume the same control input as in the previous systems, add white Gaussian process noise to $[\dot{x}, \dot{y}]$, and add white Gaussian measurement noise to position $[x, y]$. 
\Ignore{
\begin{equation}
\begin{aligned}
&\mathbf{A} =
\begin{bmatrix}
0 & 0 & 1 & 0 \\
0 & 0 & 0 & 1 \\
0 & 0 & 0 & 0 \\
0 & 0 & 0 & 0 \\
\end{bmatrix} \ \ 
\mathbf{G} =
\begin{bmatrix}
0 \\
0 \\
1 \\
1  
\end{bmatrix} \ \
\HM{} =
\begin{bmatrix}
1 & 0\\
0 & 1\\
0 & 0\\
0 & 0\\
\end{bmatrix}^T \\ 
&\mathbf{\Gamma} = 
\begin{bmatrix}
0 & 0\\ 
0 & 0\\
1 & 0\\
0 & 1
\end{bmatrix} \ \
\mathbf{V} =
\begin{bmatrix}
v_0 & 0\\
0 & v_1
\end{bmatrix} \ \
\mathbf{W} =
\begin{bmatrix}
w_0 & 0\\
0 & w_1
\end{bmatrix}
\end{aligned}
\end{equation},
where $v_0 = 1, v_1 = 2$ and $w_0= 0.2, w_1=0.1$.  
Applying eq. \eqref{eq:VanLoans}, the discrete time system is}
The discrete time system is
\begin{equation}
\begin{aligned}
    &\mathbf{F} = 
    \begin{bmatrix}
    1 & 0 & \Delta t & 0\\
    0 & 1 & 0 & \Delta t\\
    0 & 0 & 1 & 0\\
    0 & 0 & 0 & 1\\
    \end{bmatrix}
    \ \
    \mathbf{B} = 
    \begin{bmatrix}
    0.5\Delta t^2 \\ 0.5\Delta t^2 \\ \Delta t \\ \Delta t
    \end{bmatrix}
    \\
    &\mathbf{Q} = 
    \begin{bmatrix}
     \frac{\Delta t^3}{3}V_0 & 0 & \frac{\Delta t^2}{2}V_0 & 0\\
     0 & \frac{\Delta t^3}{3}V_1 & 0 & \frac{\Delta t^2}{2}V_1\\
     \frac{\Delta t^2}{2}V_0 & 0 & \Delta t V_0 & 0\\
     0 & \frac{\Delta t^2}{2}V_1 & 0 & \Delta t V_1\\
    \end{bmatrix}
\end{aligned}
\end{equation}
We apply the same optimization methods as in the tracking 1D example for 50 independent trials. We need to increase the GPBO initial sample to 120 and the iteration to 300 since the dimension is higher.\\
\indent For each optimization result of the algorithm, we apply it to the Kalman Filter again with 200 Monte Carlo runs and record the $\E{\avgnis{k}}, \E{\avgnees{k}}, \E{\avgnis{k}\avgnis{k}}, \E{\avgnees{k}\avgnees{k}}$ for validation. We choose sample time $dt = 0.1$ to collect the data. Note that we don't draw the box plot of the downhill sample algorithm with a single sample time since its value is too large. The range of the box plots of other methods will be too small to visualize if we draw it. As we can see, generally the proposed method has better NIS and covariance. However, we still hope them can have a better match to the expectations, which brings questions to our future work. 1: If we use NIS based cost function, what the NEES value will be from the optimization result? 2: Is it possible to add the covariance into the cost function constraints?  If so, can the NEES/NIS variance value be more consistent with the expectation?\\
\indent Finally, we perform the direct consistency check of the proposed method for both 1D and 2D system. We randomly choose one the optimization result and apply it to the Kalman filter. Then we plot each timestep's error and the 2$\sigma$ boundary, where the $\sigma = \sqrt{\covCond{k}{k}}$. If the system is consistent, around 95$\%$ error should be within $2\sigma$ range.

\section{Conclusion}
\label{sct:conclusion}

%conclusion.tex
%

We have demonstrated that there is implicit dependence of $\Jnees$ on $\Delta{t}$, and that as a result, many auto-tuning algorithms face significant challenge short of running a search over multi-dimensional space for optimal noise parameters and their corresponding $\Delta{t}$. While it is true that around the groundtruth noise parameters, $\Jnees$ will be small independent of what $\Delta{t}$ is, we identify that for other guesses at noise parameters, the $\Jnees$ is highly dependent on timestep choice. To address this, we propose a simple sampling procedure that appears to remedy this problem while allaying grievous increases in computational cost. Finally, we demonstrate this new approach on an auto-tuning algorithm for Kalman filter noise parameters. As future work, we believe a proof of this technique would be highly valuable. Furthermore, there exists an open investigation into the effectiveness of various statistical tests for significance in the mean and variance of the auto-tuning algorithms.

%\nra{needs to be updated!!!}
%As a black box optimization method, \BO{} simplifies what we need to know about a system to get the minimum cost. We used two examples to show that this algorithm can be useful. This novel approach can also help the practitioners get the optimal noise covariance much faster than tuning the KF manually.\\
%\indent In this work, we firstly observe that, in Van Loan's method, only around the groundtruth the optimized noise parameter can have small cost on different sample time. Based on this, we novely design a cost function with different sample time to force the final result converge to the groundtruth.\\
%\indent In the future, we plan to provide a closed form prove for the current phenomenon. At the same time, we aim to apply and test NIS based cost function and add the $\chi^2$ covariance into the cost function. Finally, we'd like to implement it to EKF and real system to observe the performance. \\

\appendix

\section{Theoretical Calculation of the NEES}
\label{adx:nees}

In this appendix, we compute the expression to derive a closed-form solution for the NEES directly from the system equations. We assume that the system model equations $\mathbf{A}_t$, $\mathbf{G}_t$, $\boldsymbol{\Gamma}_t$ and $\HM{t}$ are correct. Only the noise intensities are unknown. For simplicity, we follow the work of Nishimura and Hellner and compute the NEES of the \emph{predicted\/} covariance.

First consider the filter which has been tuned with the intensities $\V$ and $\W$. Using van Loan's method, we compute the discrete time process model together with the noise covariance matrices $\Q{k}(\V)$ and $\R{k}(\W)$, where we have included the intensities to emphasise the functional dependency. The filter will then predict the covariance history according to
\begin{equation}
\begin{split}
\covCond{\kCur}{\kLst}(\V,\W)&=\mathbf{X}_{\kCur}(\V,\W)\covCond{\kLst}{k-2}(\V,\W)\mathbf{X}_{\kCur}^\top(\V,\W)\\
&\quad
+\KK{\kCur}(\V,\W)\R{k}(\W)\KKt{\kCur}(\V,\W)\\
&\quad+\Q{k}(\V),
\end{split}
\label{eqn:pred_nishimura}
\end{equation}
where
\begin{align}
\KK{\kCur}(\V,\W)&=\F{\kCur}\Kw{\kCur}(\V,\W)\\
\mathbf{X}_{\kCur}(\V,\W)&=\F{\kCur}-\KK{\kCur}(\V,\W)\HM{\kCur},
\end{align}
and $\Kw{\kCur}(\V,\W)$ is the usual Kalman filter weight.

However, the real system has noise intensities $\V^a$ and $\W^a$. Given that there are no errors in the system model equations, the expected value of the mean squared error of the filter is actually
\begin{equation}
\begin{split}
&\covCondTrue{\kCur}{\kLst}(\V,\W,\V^a,\W^a)=\\
&\quad\mathbf{X}_{\kCur}(\V,\W)\covCondTrue{\kLst}{k-2}(\V,\W,\V^a,\W^a)\mathbf{X}_{\kCur}^\top(\V,\W)\\
&\quad
+\KK{\kCur}(\V,\W)\R{k}(\W^a)\KKt{\kCur}(\V,\W)+\Q{k}(\V^a).
\end{split}
\end{equation}

Given this, the expected value of the NEES is
\begin{equation}
\begin{split}
&\E{\epsilon_{\mathbf{x},k}}(\V,\W,\V^a,\W^a)=\\
&\qquad\mathrm{trace}\left(\covCondi{\kCur}{\kLst}(\V,\W)\covCondTrue{\kCur}{\kLst}(\V,\W,\V^a,\W^a)\right).
\end{split}
\end{equation}

The $J_{NEES}$ of this value is
\begin{equation}
J_{NEES}(\V,\W,\V^a,\W^a)=\left|\log\frac{\E{\epsilon_{\mathbf{x},k}}(\V,\W,\V^a,\W^a)}{n_x}\right|.
\label{eqn:theoretical_jnees}
\end{equation}



\begin{thebibliography}{10}
\providecommand{\url}[1]{#1}
\csname url@samestyle\endcsname
\providecommand{\newblock}{\relax}
\providecommand{\bibinfo}[2]{#2}
\providecommand{\BIBentrySTDinterwordspacing}{\spaceskip=0pt\relax}
\providecommand{\BIBentryALTinterwordstretchfactor}{4}
\providecommand{\BIBentryALTinterwordspacing}{\spaceskip=\fontdimen2\font plus
\BIBentryALTinterwordstretchfactor\fontdimen3\font minus
  \fontdimen4\font\relax}
\providecommand{\BIBforeignlanguage}[2]{{%
\expandafter\ifx\csname l@#1\endcsname\relax
\typeout{** WARNING: IEEEtran.bst: No hyphenation pattern has been}%
\typeout{** loaded for the language `#1'. Using the pattern for}%
\typeout{** the default language instead.}%
\else
\language=\csname l@#1\endcsname
\fi
#2}}
\providecommand{\BIBdecl}{\relax}
\BIBdecl

\bibitem{aakesson2007tool}
B.~M. {\AA}kesson, J.~B. J{\o}rgensen, N.~K. Poulsen, and S.~B. J{\o}rgensen,
  ``A tool for {K}alman filter tuning,'' in \emph{Computer Aided Chemical
  Engineering}.\hskip 1em plus 0.5em minus 0.4em\relax Elsevier, 2007, vol.~24,
  pp. 859--864.

\bibitem{AKESSON2008769}
B.~M. Åkesson, J.~B. Jørgensen, N.~K. Poulsen, and S.~B. Jørgensen, ``A
  generalized autocovariance least-squares method for {K}alman filter tuning,''
  \emph{Journal of Process Control}, vol.~18, no.~7, pp. 769--779, 2008.

\bibitem{powell2002automated}
T.~D. Powell, ``Automated tuning of an extended {K}alman filter using the
  downhill simplex algorithm,'' \emph{Journal of Guidance, Control, and
  Dynamics}, vol.~25, no.~5, pp. 901--908, 2002.

\bibitem{chen2018weak}
Z.~Chen, C.~Heckman, S.~Julier, and N.~Ahmed, ``Weak in the nees?: Auto-tuning
  {K}alman filters with {B}ayesian optimization,'' in \emph{2018 21st
  International Conference on Information Fusion (FUSION)}.\hskip 1em plus
  0.5em minus 0.4em\relax IEEE, 2018, pp. 1072--1079.

\bibitem{mu2009automatic}
T.~Mu and A.~K. Nandi, ``Automatic tuning of l2-svm parameters employing the
  extended kalman filter,'' \emph{Expert Systems}, vol.~26, no.~2, pp.
  160--175, 2009.

\bibitem{scardua2016automatic}
L.~A. Scardua and J.~J. Da~Cruz, ``Automatic tuning of the unscented {K}alman
  filter and the blind tricyclist problem: an optimization problem,''
  \emph{IEEE Control Systems Magazine}, vol.~36, no.~3, pp. 70--85, 2016.

\bibitem{mohindergrewalVanLoanMethod2015}
{Mohinder Grewal} and A.~Andrews, ``Van {{Loan}}'s {{Method}} for {{Computing}}
  ${{Q}}\_k$ from {{Continuous}} ${{Q}}$,'' in \emph{{K}alman {{Filtering}}:
  {{Theory}} and {{Practice}} with {{MATLAB}}}, 2015, pp. 150--152.

\bibitem{Kalman-JBE-1961}
R.~E. Kalman and R.~S. Bucy, ``New results in linear filtering and prediction
  theory,'' \emph{Journal of Basic Engineering}, vol.~83, no.~1, pp. 95--108,
  1961.

\bibitem{Bar-Shalom2001}
Y.~Bar-Shalom, X.~Li, and T.Kirubarajan, \emph{{Estimation with Applications to
  Navigation and Tracking}}.\hskip 1em plus 0.5em minus 0.4em\relax New York:
  Wiley, 2001.

\bibitem{zhangIdentificationNoiseCovariances2020}
L.~Zhang, D.~Sidoti, A.~Bienkowski, K.~R. Pattipati, Y.~{Bar-Shalom}, and D.~L.
  Kleinman, ``On the {{Identification}} of {{Noise Covariances}} and {{Adaptive
  {K}alman Filtering}}: {{A New Look}} at a 50 {{Year}}-{{Old Problem}},''
  \emph{IEEE Access}, vol.~8, pp. 59\,362--59\,388, 2020.

\bibitem{dunik2017noise}
J.~Dun{\'\i}k, O.~Straka, O.~Kost, and J.~Havl{\'\i}k, ``Noise covariance
  matrices in state-space models: A survey and comparison of estimation
  methods—{P}art {I},'' \emph{International Journal of Adaptive Control and
  Signal Processing}, vol.~31, no.~11, pp. 1505--1543, 2017.

\bibitem{Bishop2006}
C.~M. Bishop, \emph{{Pattern recognition and machine learning}}.\hskip 1em plus
  0.5em minus 0.4em\relax New York: Springer, 2006.

\bibitem{barratt2020fitting}
S.~T. Barratt and S.~P. Boyd, ``Fitting a {K}alman smoother to data,'' in
  \emph{2020 American Control Conference (ACC)}.\hskip 1em plus 0.5em minus
  0.4em\relax IEEE, 2020, pp. 1526--1531.

\bibitem{dunik2020covariance}
J.~Dun{\'\i}k, O.~Kost, O.~Straka, and E.~Blasch, ``Covariance estimation and
  {G}aussianity assessment for state and measurement noise,'' \emph{Journal of
  Guidance, Control, and Dynamics}, vol.~43, no.~1, pp. 132--139, 2020.

\bibitem{ko2009gp}
J.~Ko and D.~Fox, ``{GP-B}ayes filters: {B}ayesian filtering using gaussian
  process prediction and observation models,'' \emph{Autonomous Robots},
  vol.~27, no.~1, pp. 75--90, 2009.

\bibitem{oshman2000optimal}
Y.~Oshman and I.~Shaviv, ``Optimal tuning of a {K}alman filter using genetic
  algorithms,'' in \emph{AIAA Guidance, Navigation, and Control Conference and
  Exhibit}, 2000, p. 4558.

\bibitem{saha2013robustness}
M.~Saha, R.~Ghosh, and B.~Goswami, ``Robustness and sensitivity metrics for
  tuning the extended {K}alman filter,'' \emph{IEEE Transactions on
  Instrumentation and Measurement}, vol.~63, no.~4, pp. 964--971, 2013.

\bibitem{piche2016online}
R.~Pich{\'e}, ``Online tests of {K}alman filter consistency,''
  \emph{International Journal of Adaptive Control and Signal Processing},
  vol.~30, no.~1, pp. 115--124, 2016.

\bibitem{gibbs2013new}
R.~G. Gibbs, ``New {K}alman filter and smoother consistency tests,''
  \emph{Automatica}, vol.~49, no.~10, pp. 3141--3144, 2013.

\bibitem{minasny2005matern}
B.~Minasny and A.~B. McBratney, ``The mat{\'e}rn function as a general model
  for soil variograms,'' \emph{Geoderma}, vol. 128, no. 3-4, pp. 192--207,
  2005.

\bibitem{JMLR:v15:martinezcantin14a}
\BIBentryALTinterwordspacing
R.~Martinez-Cantin, ``Bayesopt: A {B}ayesian optimization library for nonlinear
  optimization, experimental design and bandits,'' \emph{Journal of Machine
  Learning Research}, vol.~15, pp. 3915--3919, 2014. [Online]. Available:
  \url{http://jmlr.org/papers/v15/martinezcantin14a.html}
\BIBentrySTDinterwordspacing

\end{thebibliography}
\end{document}